\pdfoutput=1
\documentclass[11pt,twoside,a4paper,cmspaper,final,collab]{cms-tdr}
\def\svnVersion{c35e27e-D}\def\svnDate{2025/07/23}\def\cmsCernNoTag{CERN-EP-2025-156}\def\cmsCernDate{\today}\def\cmsMessage{Submitted to the Journal of High Energy Physics}
\begin{document}\cmsNoteHeader{EXO-24-038}

\providecommand{\cmsLeft}{left\xspace}
\providecommand{\cmsRight}{right\xspace}
\newcommand{\PY}{\ensuremath{\mathrm{Y}}\xspace}
\newcommand{\PS}{\ensuremath{\mathrm{S}}\xspace}
\newcommand{\PSuu}{\ensuremath{\PS_{\PQu\PQu}}\xspace}
\newcommand{\PSdd}{\ensuremath{\PS_{\PQd\PQd}}\xspace}
\newcommand{\mjj}{\ensuremath{\overline{m}_{\mathrm{2j}}}\xspace}
\newcommand{\mjjjj}{\ensuremath{m_{\mathrm{4j}}}\xspace}
\newcommand{\atrue}{\ensuremath{\alpha_{\mathrm{true}}}\xspace}
\newcommand{\yuu}{\ensuremath{y_{\PQu\PQu}}\xspace}
\newcommand{\ydd}{\ensuremath{y_{\PQd\PQd}}\xspace}
\newcommand{\ychi}{\ensuremath{y_{\PGc}}\xspace}
\newcommand{\yomega}{\ensuremath{y_{\PGo}}\xspace}

\cmsNoteHeader{EXO-24-038}
\title{Search for resonant production of pairs of dijet resonances through broad mediators in proton-proton collisions at \texorpdfstring{$\sqrt{s} = 13\TeV$}{sqrt(s) = 13 TeV}}

\author*[cern]{The CMS Collaboration}

\date{\today}

\abstract{
A reinterpretation of a prior narrow-resonance search is performed to investigate the resonant production of pairs of dijet resonances via broad mediators. This analysis targets events with four resolved jets, requiring dijet invariant masses greater than 0.2\TeV and four-jet invariant masses greater than 1.6\TeV. The search uses a data sample corresponding to an integrated luminosity of 138\fbinv collected by the CMS experiment in proton-proton collisions at $\sqrt{s} = 13\TeV$. The reinterpretation considers the production of new heavy four-jet resonances, with widths ranging from 1.5 to 10\% of their mass, which decay to a pair of dijet resonances. This analysis probes resonant production in the four-jet and dijet mass distributions. Upper limits at 95\% confidence level and significances are reported on the production cross section of new resonances as functions of their widths and masses, between 2 and 10\TeV. In particular, at a four-jet resonance mass of 8.6\TeV, the local (global) significance ranges from 3.9 (1.6) to 3.6 (1.4) standard deviations (s.d.) as the resonance width is increased from 1.5 to 10\%. This relative insensitivity to the choice of width indicates that a broad resonance is an equally valid interpretation of this excess. The broad resonance hypothesis at a resonance mass of 8.6\TeV is supported by the presence of an event with a four-jet mass of 5.8\TeV and an average dijet mass of 2.0\TeV. Also, we report the reinterpretation of a second effect, at a four-jet resonance mass of 3.6\TeV, which has a local (global) significance of up to 3.9 (2.2) s.d.
}

\hypersetup{%
pdfauthor={CMS Collaboration},%
pdftitle={Search for resonant production of pairs of dijet resonances through broad mediators in proton-proton collisions at sqrt(s) = 13TeV},%
pdfsubject={CMS},%
pdfkeywords={CMS, paired dijets}} 
\maketitle 

\section{Introduction and motivation}
\label{sec:IntroductionAndMotivation}

Measurements of pairs of dijet resonances with the same invariant mass are powerful probes of physics beyond the standard model (SM). The pairs of dijet resonances can be  produced resonantly via a massive $s$-channel mediator, which can be broad.

Generic searches for resonant production of pairs of dijet resonances, where each jet is individually resolved and each resonance is a particle beyond the SM, have recently been performed by the CMS~\cite{CMS-EXO-21-010} and ATLAS~\cite{PairedDijetATLAS} Collaborations. The data in both searches were used to constrain narrow mediators, with natural widths less than 0.5$\%$ of their mass, which is about an order of magnitude smaller than the experimental resolutions.

We present a reinterpretation of the Ref.~\cite{CMS-EXO-21-010} analysis based on the CERN LHC proton-proton ($\Pp\Pp$) collision data at $\sqrt{s} = 13\TeV$ collected by the CMS experiment in 2016--2018 and corresponding to an integrated luminosity of 138\fbinv, by extending it to models with resonant production of a pair of dijet resonances via a broad mediator.

\begin{figure}[hbtp]
\centering
\includegraphics[width=0.70\textwidth]{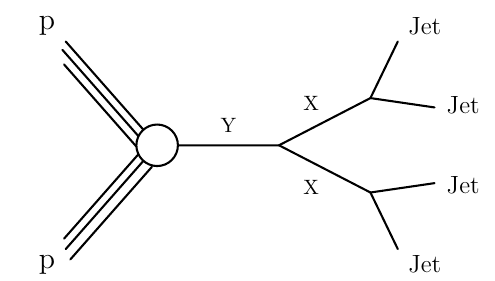}
\caption{Resonant production via a particle, \PY, of pairs of dijet resonances, \PX.} 
\label{fig:Feynman} 
\end{figure}

As shown in Fig.~\ref{fig:Feynman}, we consider resonant production of pairs of dijet resonances,
\begin{equation} 
   \Pp\Pp \to \PY \to \PX\PX \to (\mathrm{jj})(\mathrm{jj}),
   \label{eq:first}
\end{equation}
where the intermediate state is a massive ($>$2.0\TeV) new particle, \PY, decaying to identical dijet resonances, \PX, where both jets, $\mathrm{j}$, within each dijet resonance are individually reconstructed. The natural width of particle \PY is considered to be comparable or larger than the experimental resolution, thus forming a broad resonance, whereas particle \PX is characterized by a narrow intrinsic width. 

As a benchmark, we consider a diquark model~\cite{Robertpaper}, where the intermediate state is a color sextet diquark, \PSuu or \PSdd, which carries an electric charge of $4/3$ or $-2/3$, respectively, and is produced from the annihilation of two up (\PQu) or down (\PQd) quarks. The dijet resonance is a vector-like quark, \PGc or \PGo, that decays at one loop into a gluon (\Pg) and an up or down quark, respectively:
\begin{equation}
\begin{aligned}
   \PQu\PQu &\to \PSuu \to \PGc\PGc \to (\PQu\Pg)(\PQu\Pg)\\
  \PQd\PQd &\to \PSdd \to \PGo\PGo \to (\PQd\Pg)(\PQd\Pg).
\label{eq:Diquarks}
\end{aligned}
\end{equation}
The scalar diquark is a suitable benchmark because it is produced with a large cross section, due to the high probability of finding valence quarks with large fractional momenta within the proton. Representatively, for an \PSuu diquark with a mass of 8.4 (3.6)\TeV decaying to vector-like quarks each with a mass of 2.1 (1.0)\TeV, the cross section is $7.9\times 10^{-5}$ ($3.7\times 10^{-1}$)\unit{pb} for a width of 1.5\% and $9.5\times 10^{-4}$ (2.3)\unit{pb} for a width of 10\%. The phenomenology of the diquark model and its implications for CMS searches with dijets has been explored in Refs.~\cite{Robertpaper, dobrescu2024tevscaleparticleslhcevents, dobrescu2019lhcprobes10tev, hxll-41ly}.

Throughout this paper, the mass of both \PSuu and \PSdd is denoted by $M_{\PS}$. The lowercase $m$ denotes reconstructed dijet and four-jet masses, and the uppercase $M$ denotes true resonance masses. Furthermore, we will quote the resonance width as a percentage of the resonance mass, and denote it as the ratio of natural width ($\Gamma$) to resonance mass ($M$).

This reinterpretation analysis is motivated by the data recorded with the CMS and ATLAS experiments, presented in Figs. 2 and 3(a) of Refs.~\cite{CMS-EXO-21-010} and~\cite{PairedDijetATLAS}, respectively. Two events were observed in the CMS search with a four-jet mass, \mjjjj, of 8\TeV and an average dijet mass of the two dijet pairs, \mjj, of 2\TeV. These events were interpreted as potential candidates for a narrow resonance, with a local significance of 3.9 standard deviations (s.d.) and a global significance of 1.6 s.d. An additional CMS event with $\mjjjj = 5.8\TeV$ and $\mjj =2.0\TeV$ is found within the reported 95\% probability contour of a narrow resonance simulation, with a diquark mass of 8.4\TeV and vector-like quark mass of 2.1\TeV~\cite{CMS-EXO-21-010}. Furthermore, an ATLAS event with $\mjjjj = 6.6\TeV$ and $\mjj = 2.2\TeV$ is also found within the same contour. The similar values of \mjj of these events, all around 2\TeV, and the wider spread of the observed \mjjjj, make them potential candidates for a broad four-jet resonance decaying into a pair of narrow dijet resonances. This is a clear motivation to perform a reinterpretation of the previous CMS search, to explore the scenario where the mediator \PY is broad and the resonance \PX is narrow, and report the corresponding limits at 95\% confidence level (\CL) and significance for that hypothesis.

Tabulated results are provided in the HEPData record for this analysis~\cite{HEPData}.

\section{Simulated data samples}
\label{sec:SimulatedDataSamples}

A background sample of quantum chromodynamics (QCD) multijet events is produced, for the optimization of the search and for qualitative comparisons with the observed data and the simulated signal data sets, as detailed in Ref.~\cite{CMS-EXO-21-010}.

The benchmark model for the search is based on the aforementioned model of a diquark decaying to pairs of vector-like quarks, defined in Eq.~\eqref{eq:Diquarks}. The \MGvATNLO~2.6.5~\cite{madgraph} generator, with additional code specifying the model~\cite{Robertpaper}, is used to generate these events and \textsc{MadSpin} is used to simulate the decays of the vector-like quarks to quarks and gluons. Events are generated for diquark masses between 2 and 10\TeV, for vector-like quark masses between 0.22 and 4.2\TeV, and for diquark widths between 1.5 and 10\%. The width of the vector-like quarks is proportional to the square of their mass, remaining below 1.3\% for masses below 2.5\TeV and reaching up to 3.7\% at a mass of 4.2\TeV.

For the benchmark signal simulations, \PYTHIA~8.205~\cite{Sjostrand:2014zea} is used to produce the parton shower and the resulting final-state particles. The CP5 underlying event tune~\cite{Khachatryan:2015pea} is used with the NNPDF3.1 PDF set~\cite{Ball:2014uwa} at next-to-next-to-leading order.

The effects of the CMS detector, which is described in detail in Ref.~\cite{refCMS}, are modeled using the \GEANTfour toolkit~\cite{geant4}. Additional $\Pp\Pp$ interactions within the same or neighboring bunch crossings (pileup) are taken into account by adding simulated minimum bias interactions to all the generated event samples. The pileup distribution is weighted to match that observed in the data.

\section{Signal simulation, data, and QCD}
\label{sec:Signal}

\begin{figure}[bp]
\centering
\includegraphics[width=0.49\textwidth]{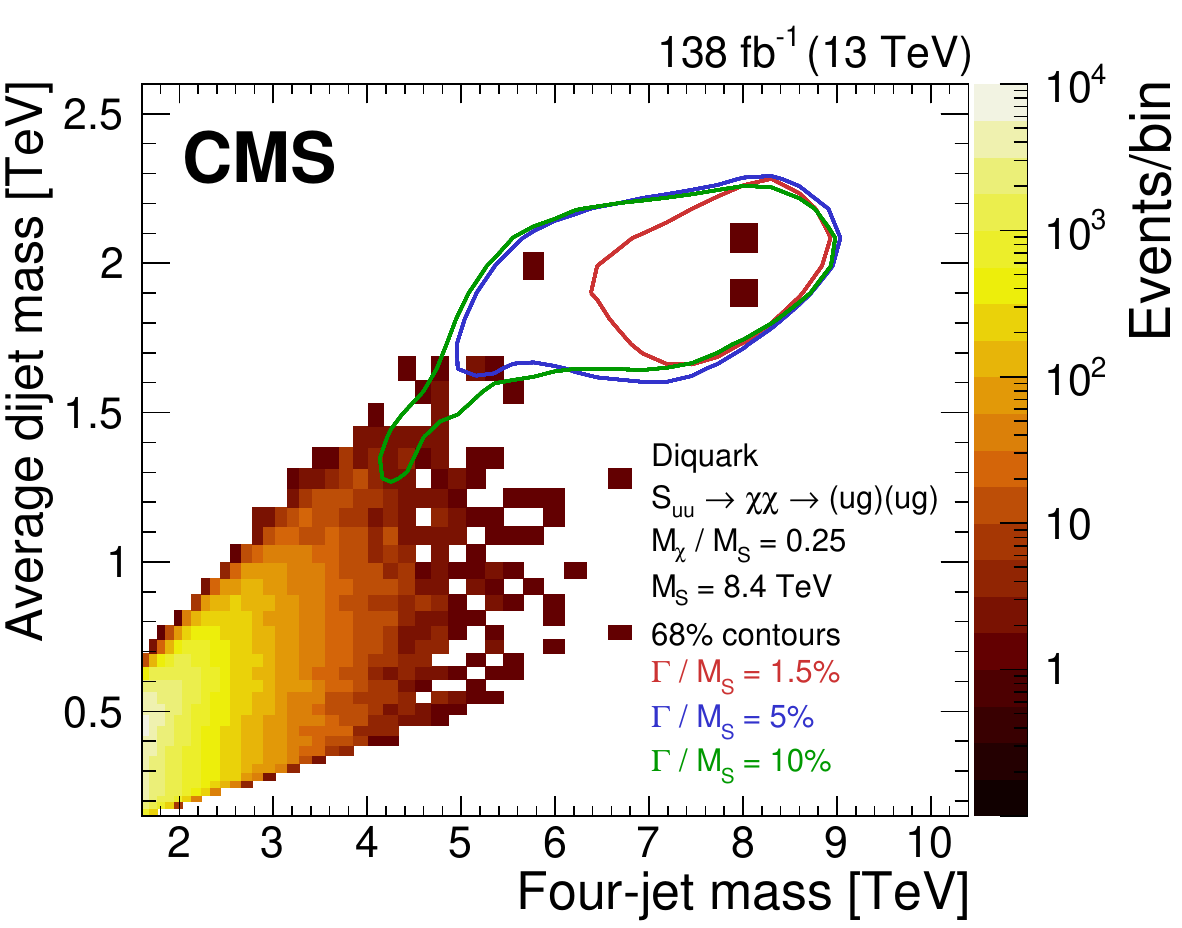} 
\includegraphics[width=0.49\textwidth]{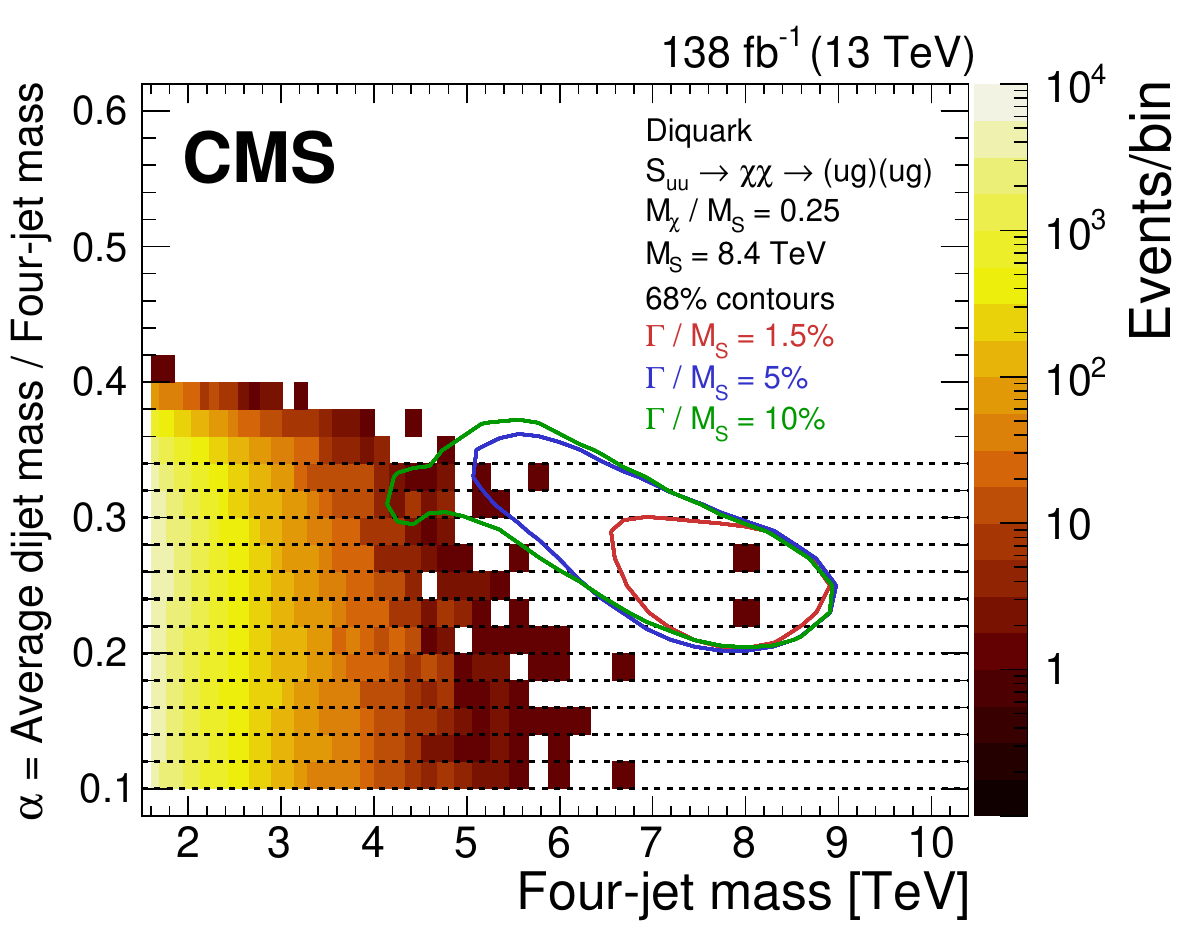} \\ 
\includegraphics[width=0.49\textwidth]{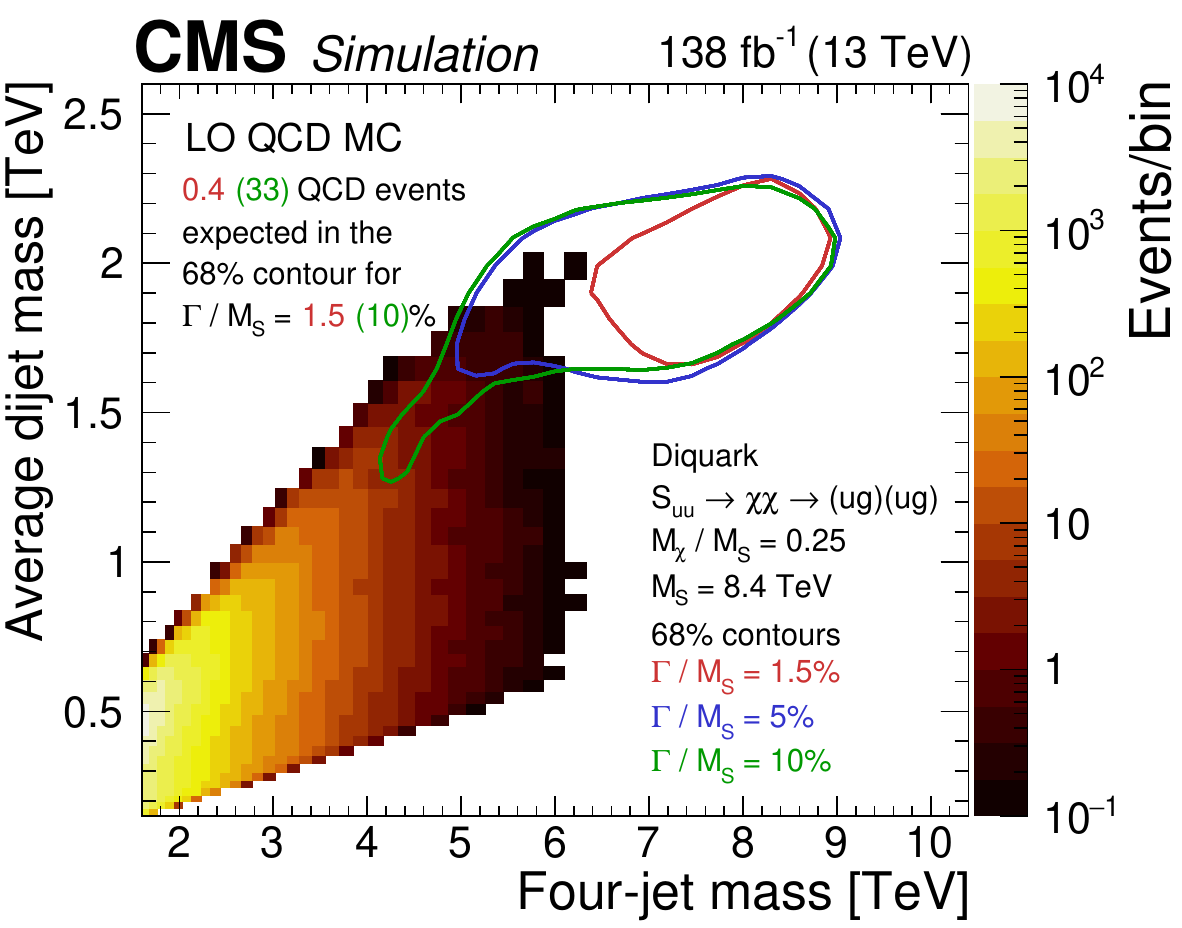} 
\includegraphics[width=0.49\textwidth]{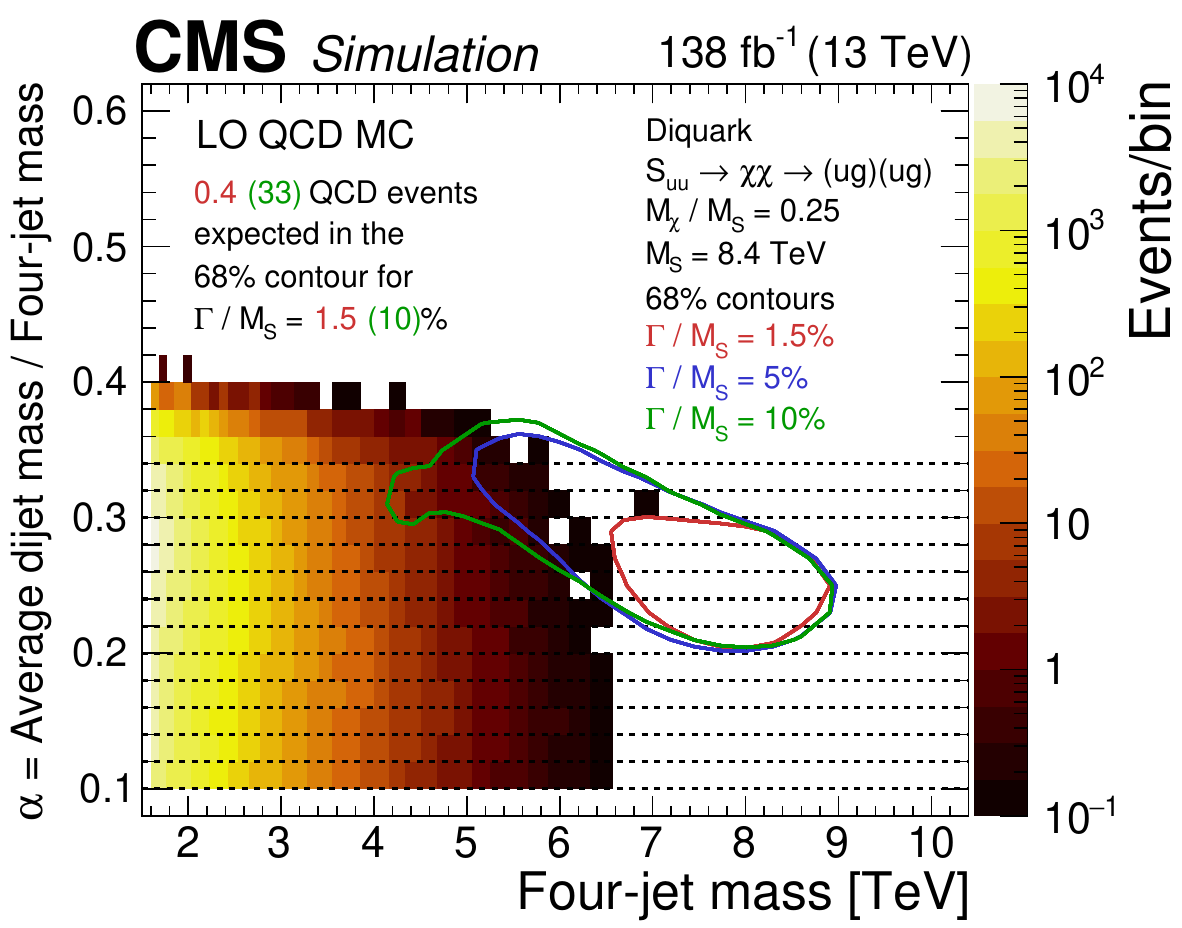}
\caption{Observed number of events (upper) and predictions of a leading order (LO) QCD simulation (lower), normalized to an integrated luminosity of 138\fbinv. The number of events (color scale) are examined within bins of the four-jet mass and the average mass of the two dijets (\cmsLeft) and within bins of \mjjjj and the ratio $\alpha$ (\cmsRight), which is \mjj divided by \mjjjj. The solid curves show the 68\% probability contours from a signal simulation of a diquark with a mass of 8.4\TeV, decaying to a pair of vector-like quarks, each with a mass of 2.1\TeV. The red, blue, and green probability contours correspond to 1.5, 5, and 10\% diquark widths, respectively. The \cmsRight plots also show the thirteen $\alpha$ bins used to define the \mjjjj distributions (dashed lines). The number of SM events, predicted by the LO QCD simulation, enclosed within the 68\% probability contours of signals with 1.5 and 10\% widths are 0.4 and 33, respectively.} 
\label{fig:2D_Data_QCD} 
\end{figure}

The event reconstruction, selection, and jet pairing algorithm are identical to those of Ref.~\cite{CMS-EXO-21-010}. The analysis strategy is to simulate the signal shapes of broad resonances, and perform a signal plus background fit of the data, exactly as in Ref.~\cite{CMS-EXO-21-010}. Only the signal shapes are new. Figure~\ref{fig:2D_Data_QCD} (upper-left) shows the two-dimensional (2D) distribution of the observed number of events as a function of \mjj and \mjjjj, using data from Ref.~\cite{CMS-EXO-21-010}, but compared here to the 68\% probability contours of simulated resonances with new widths of 1.5, 5, and 10\%. The wider contours, for resonances of widths 5 and 10\%, completely contain all three candidate events with \mjj around 2\TeV. The tail-like structure observed in the contour for the 10\% width arises from incorrect jet combinations assigned by the pairing algorithm.

As in Ref.~\cite{CMS-EXO-21-010}, the actual fit is conducted in bins of the variable $\alpha = \mjj/\mjjjj$. Figure~\ref{fig:2D_Data_QCD} (upper-right) shows the same data and simulated resonance contours for $\alpha$ vs. \mjjjj, along with the $\alpha$ binning that will be used for the fits. Events with $\alpha$ between 0.10 and 0.34 are considered in 12 bins of size 0.02, and events with $\alpha > 0.34$ are considered in a single bin. The bin width is chosen to be comparable to the resolution of the reconstructed $\alpha$ for narrow resonances.

Figure~\ref{fig:2D_Data_QCD} (lower) shows the 2D distribution of the number of events predicted from a QCD simulated data set, as a function of \mjj and \mjjjj (\cmsLeft), as well as $\alpha$ and \mjjjj (\cmsRight). Events originating from QCD processes are distributed with an approximately uniform density in $\alpha$, and there is a wide range of \mjjjj for each value of $\alpha$, to facilitate estimation of the background directly from the observed events, as discussed in the next section. Resonant signals, on the other hand, are observed as localized excesses in this 2D plane, as can be seen from the 68\% probability contours of resonances with a diquark mass of 8.4\TeV, depicted in Fig.~\ref{fig:2D_Data_QCD}.

The resonance width of a diquark \PSuu is equal to the sum of the partial width of its decay into a pair of up quarks,
\begin{equation}
    \frac{\Gamma(\PSuu \to \PQu\PQu)}{M_{\PS}} = \frac{\yuu^{2}}{32\pi}    
\label{eq:SuuToUpUpWidth}		
\end{equation}
and the partial width of its decay into a pair of vector-like quarks,
\begin{equation}
    \frac{\Gamma(\PSuu \to \PGc\PGc)}{M_{\PS}} = \frac{\ychi^{2}}{32\pi}(1-2\atrue^{2})(1-4\atrue^{2})^{1/2},
\label{eq:SuuToChiChiWidth}		
\end{equation}
where $\atrue = M_{\PGc} / M_{\PS}$, \yuu (\ychi) is the coupling between the diquark and up quarks (vector-like quarks), as defined in Ref.~\cite{Robertpaper}, and the benchmark value of $\beta = \yuu / \ychi$ is set to $2/3$, as in Ref.~\cite{CMS-EXO-21-010}. For each \atrue signal scenario, the couplings \yuu and \ychi are chosen accordingly so as to generate resonances with a total natural width of 1.5, 5, and 10\%, while keeping $\beta$ at its benchmark value $\beta = 2/3$. Diquark signals with a width larger than 10\% do not exhibit a peak at the resonance mass, and hence are outside the scope of this analysis. From Ref.~\cite{Robertpaper}, the branching fraction $\mathcal{B}(\PSuu \to \PGc\PGc)$ is 0.63 when the vector-like quark mass to diquark mass ratio is 0.25. More generally, the branching fraction varies from 0.68 to 0.44 as the mass ratio increases from 0.11 to 0.42. We set $\mathcal{B}(\PGc \to \PQu\Pg) = 1$, as is done in Ref.~\cite{Robertpaper}. Equations~(\ref{eq:SuuToUpUpWidth}) and~(\ref{eq:SuuToChiChiWidth}), and the above discussion of couplings, widths, branching fractions, and our benchmark values is also true for \PSdd and $\omega$ with the symbolic replacement everywhere of $\PQu\to\PQd$ and $\PGc\to\PGo$.

\begin{figure}[hbtp]
\centering
\includegraphics[width=0.70\textwidth]{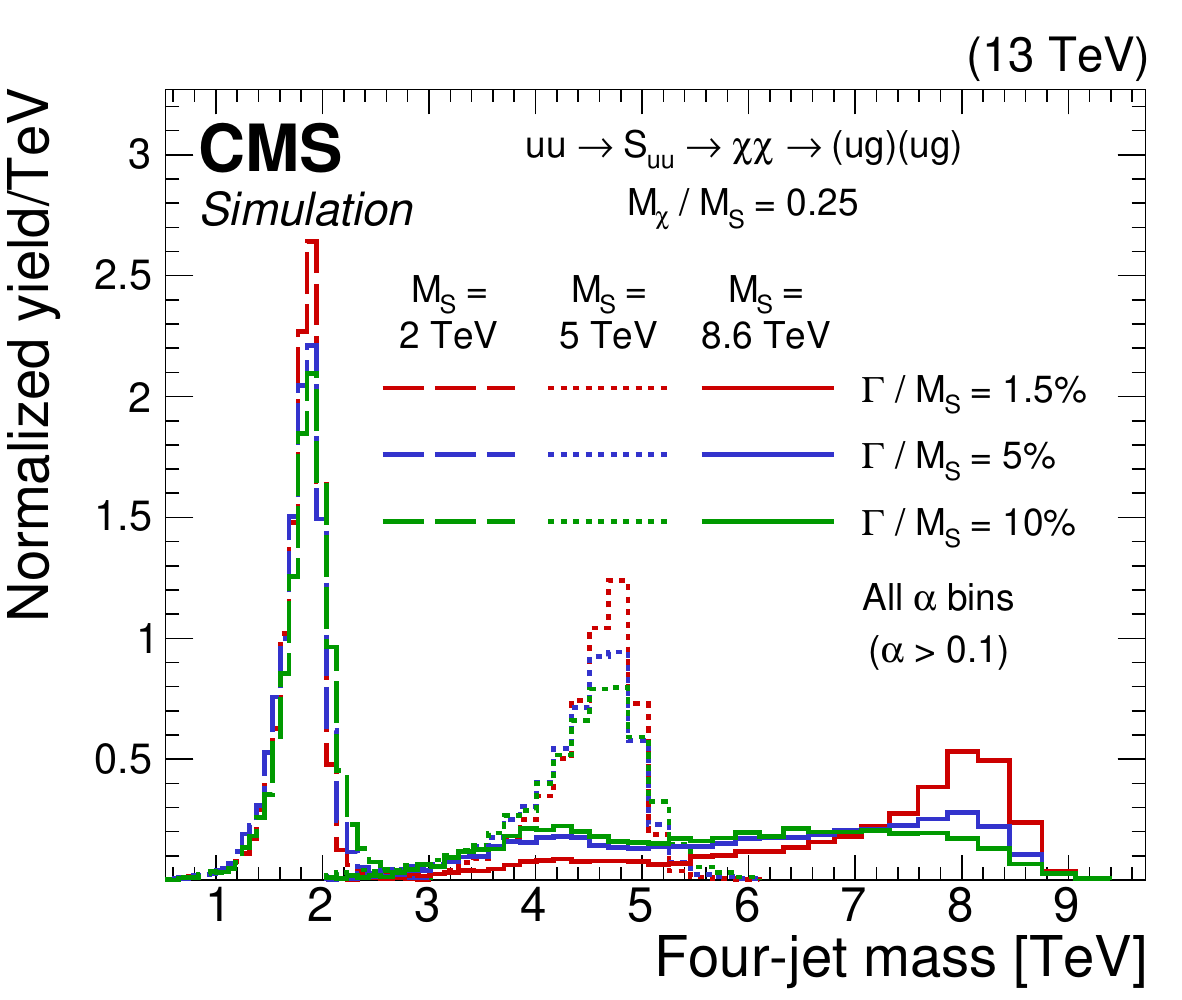}
\caption{Signal differential distributions as a function of \mjjjj for $\atrue = 0.25$, diquark masses of 2, 5, 8.6\TeV and various widths, for all $\alpha$ bins inclusively. The integral of each distribution has been normalized to unity.} 
\label{fig:SignalShapes} 
\end{figure}

Figure~\ref{fig:SignalShapes} shows the shapes of signals with $\atrue = 0.25$ and various diquark masses, across all $\alpha$ bins. Low-mass resonances exhibit a shift toward higher \mjjjj as their natural width increases, whereas intermediate-mass resonances display a broader distribution extending to both lower and higher \mjjjj. Conversely, high-mass resonances present a long, low-mass tail that becomes increasingly prominent with greater natural width, resulting in an improved description of all three candidate events with $\mjj \simeq 2\TeV$.  

\begin{figure}[hbtp]
\centering
\includegraphics[width=0.70\textwidth]{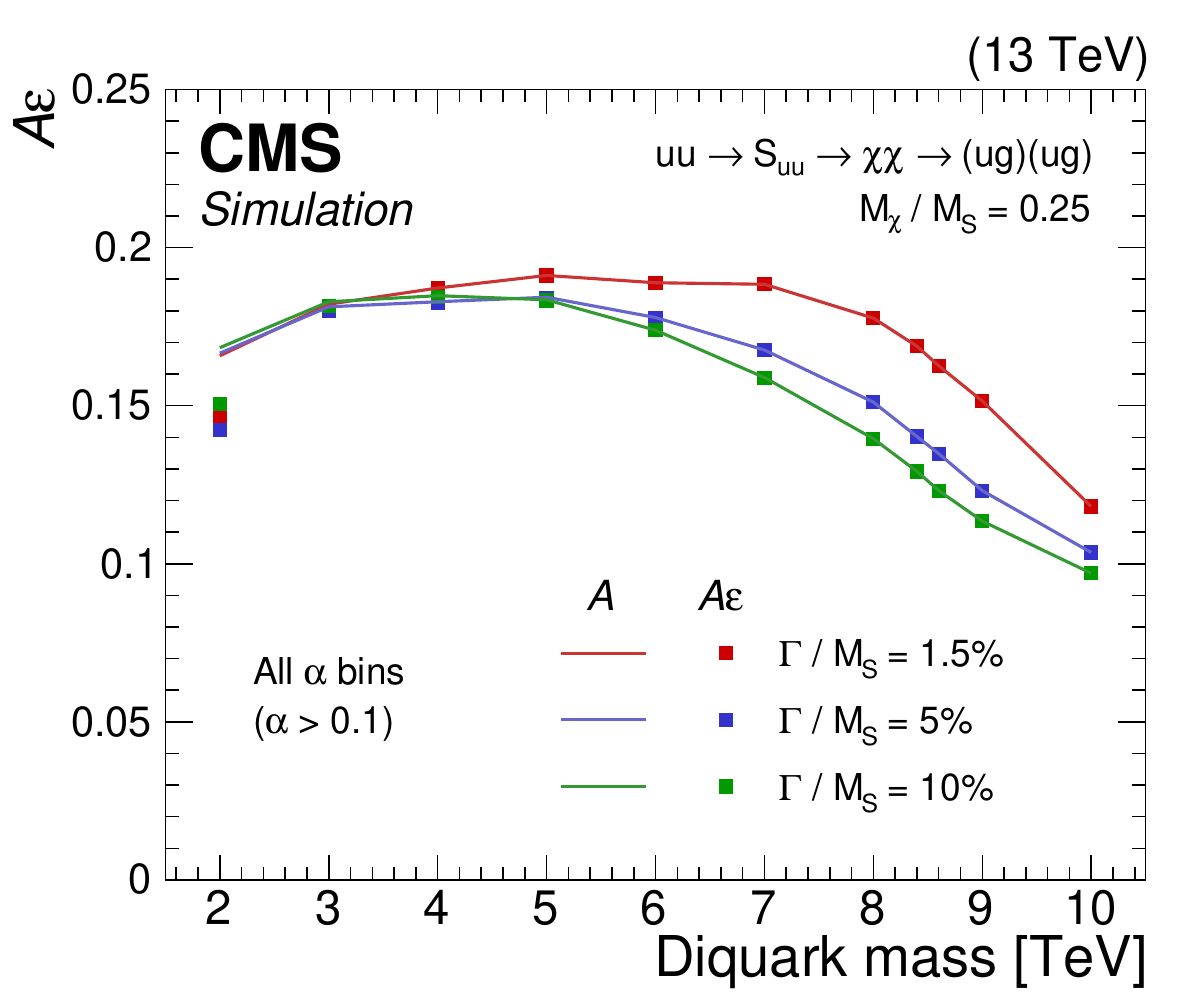}
\caption{The product of the acceptance ($A$) and efficiency of the mass selection ($\varepsilon$) for a resonant signal with $\atrue = 0.25$, shown with filled squares as a function of the diquark mass and for various diquark widths, for all $\alpha$ bins inclusively. The case when $\varepsilon=1$ is also shown as solid lines.} 
\label{fig:AcceptanceTimesEfficiency} 
\end{figure}

The products of acceptance, $A$, and efficiency, $\varepsilon$, are shown in Fig.~\ref{fig:AcceptanceTimesEfficiency}, as functions of diquark mass for all widths considered, where $A$ is defined as the fraction of generated events passing the kinematic selection criteria. In Fig.~\ref{fig:AcceptanceTimesEfficiency}, $\varepsilon$ is the fraction of signal events satisfying $\mjjjj > 1.6\TeV$, which causes the product of acceptance and efficiency to decrease at a four-jet resonance mass of 2\TeV. Figure~\ref{fig:AcceptanceTimesEfficiency} also shows the signal acceptance alone in the curves where $\varepsilon=1$. A decrease in $A$ is observed at the highest values of $M_{\PS}$, which is enhanced as the diquark width increases, and is found to originate from the mass asymmetry and $\Delta R$ requirements on the dijet pairs, as discussed in Ref.~\cite{CMS-EXO-21-010}.

\section{Background estimation}
\label{sec:BackgroundEstimation}

The challenge in searching for multi-jet resonances is to understand the background. Perturbative QCD predictions of multijet production have too many theoretical uncertainties to reliably model this background. We instead use the background modeling procedure described in Ref.~\cite{CMS-EXO-21-010}, which does not rely on simulation, but fits a parametric smooth background function to the observed data. We do this with a set of 1D distributions that span the 2D space. As in Ref.~\cite{CMS-EXO-21-010}, the search is performed in all $\alpha$ bins for $\mjjjj>1.6\TeV$ and $\mjj>0.2\TeV$.

As in previous resonance searches \cite{Sirunyan:2018xlo,Sirunyan:2016iap,Khachatryan:2016ecr,Khachatryan:2015dcf,Khachatryan:2010jd,Chatrchyan2011123,CMS:2012yf,Chatrchyan:2013qhXX,Khachatryan:2015sja,ATLAS:2015nsi,ATLAS2010,Aad:2011aj,Aad201237,ATLAS:2012pu,Aad:2014aqa,refCDFrun2}, we fit empirical functional forms to the observed multijet mass distributions. Our primary background function, which gives the best fit to the differential cross section data without including any signal component, is the modified dijet function with three parameters. As demonstrated in Ref.~\cite{CMS-EXO-21-010}, all the fits in each $\alpha$ bin adequately describe the data, with individual $p$-values~\cite{Gross_2010} ranging from 0.14 to 0.91. The combined $p$-value is 0.18 for a simultaneous fit of the background function to the data, including all thirteen $\alpha$ bins. We use two alternate background functions, the dijet function and the power-law times exponential function, both with three parameters, to estimate the systematic uncertainty in the background from variations in the functional form. This estimation employs the discrete profiling method~\cite{Dauncey:2014xga}, as detailed in Ref.~\cite{CMS-EXO-21-010}.

\begin{figure}[hbtp]
\centering
\includegraphics[width=0.99\textwidth]{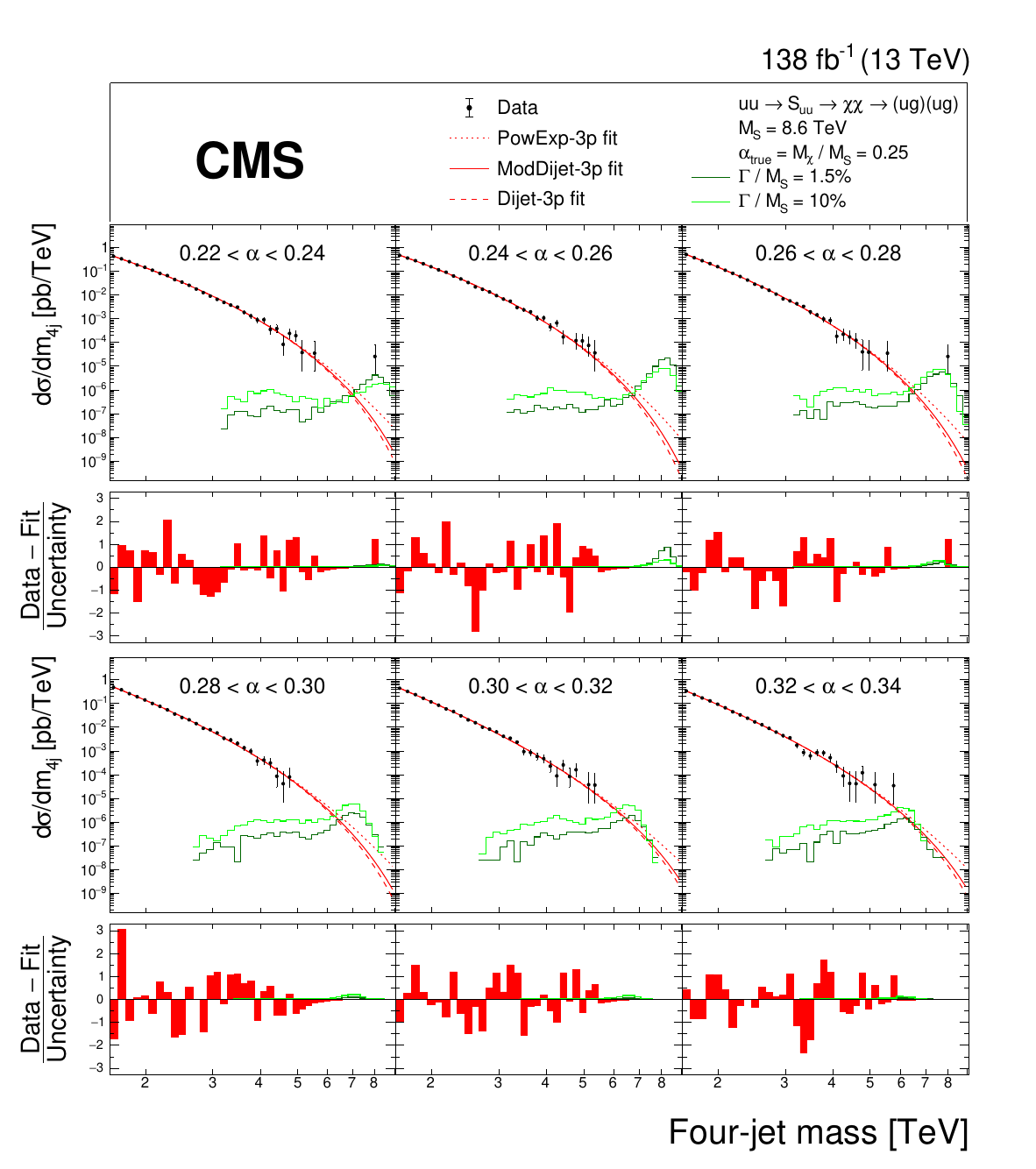}
\caption{The \mjjjj distribution in data (points), within six of the thirteen $\alpha$ bins, fitted with the background-only function, ModDijet-3p (red solid), and two alternative background functions, PowExp-3p and Dijet-3p (red dotted and dashed), with three free parameters. Examples of predicted diquark resonances with $\atrue = 0.25$, $M_{\PS} = 8.6\TeV$, and 1.5 or 10\% widths are shown, with cross sections equal to the observed upper limits at 95\% \CL. The percentage of signal across the depicted $\alpha$ bins is 90 (80)\% for the 1.5 (10)\% width. The lower panels show the pulls from the fit with the background-only function to data, calculated using the statistical uncertainty of the data.} 
\label{fig:BkgFitsWithSignalsAlphaBins_8p6} 
\end{figure}

\begin{figure}[hbtp]
\centering
\includegraphics[width=0.99\textwidth]{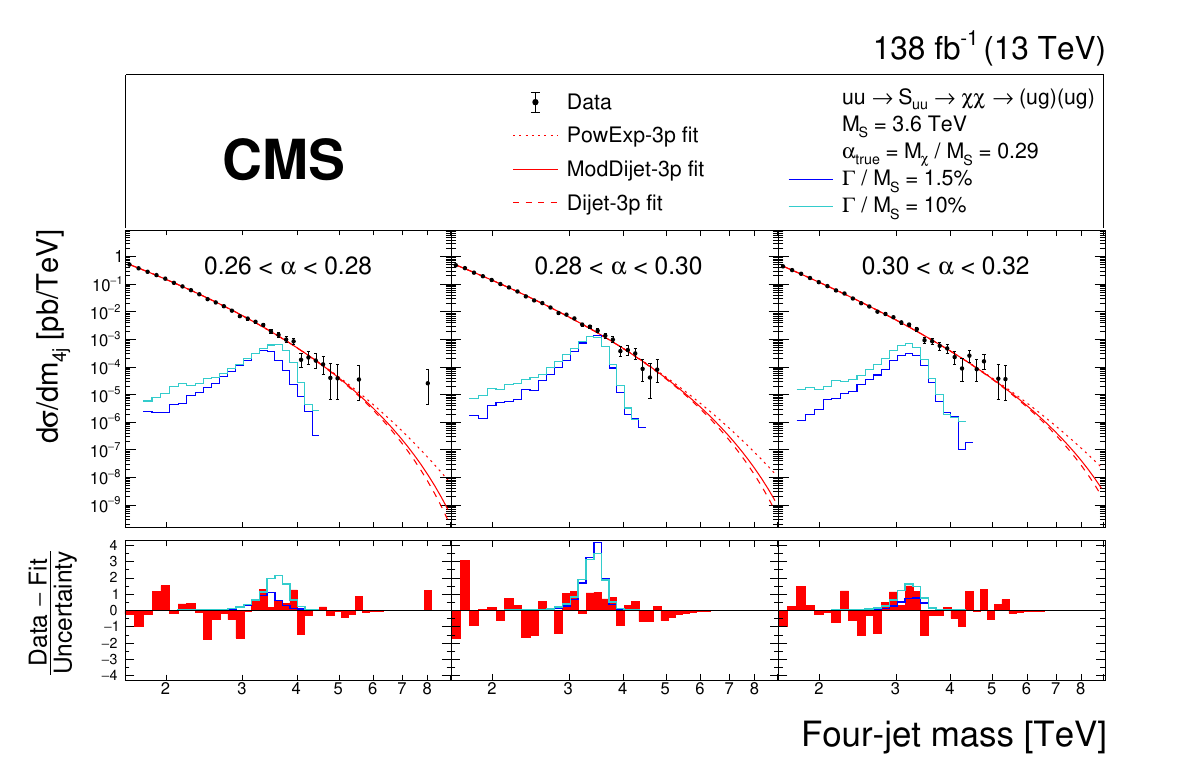}
\caption{The \mjjjj distributions of the data (points), within three of the thirteen $\alpha$ bins, fitted with the background-only function, ModDijet-3p (red solid), and two alternative background functions, PowExp-3p and Dijet-3p (red dotted and dashed), with three free parameters. Examples of predicted diquark resonances with $\atrue = 0.29$, $M_{\PS} = 3.6\TeV$, and 1.5 or 10\% widths are shown, with cross sections equal to the observed upper limits at 95\% \CL. The percentage of signal across the depicted $\alpha$ bins is 80 (70)\% for the 1.5 (10)\% width. The lower panels show the pulls from the fit with the background-only function to data, calculated using the statistical uncertainty of the data.} 
\label{fig:BkgFitsWithSignalsAlphaBins_3p6} 
\end{figure}

Figure~\ref{fig:BkgFitsWithSignalsAlphaBins_8p6} shows the observed differential cross section as a function of \mjjjj and the background fits with three functional forms, for six out of the thirteen $\alpha$ bins. Also shown are signals with $M_{\PS} = 8.6\TeV$ and $\atrue = 0.25$. When the resonance width is 1.5 (10)\%, the six $\alpha$ bins contain 90 (80)\% of the signal. Figure~\ref{fig:BkgFitsWithSignalsAlphaBins_3p6} shows the same data again for three of the $\alpha$ bins, but here compared to signals with $M_{\PS} = 3.6\TeV$ and $\atrue = 0.29$. For these
resonances, when the width is 1.5 (10)\%, the three $\alpha$ bins contain 80 (70)\% of the signal. In Section~\ref{sec:Results}, we present more information on both these signals.

\begin{figure}[hbtp]
\centering
\includegraphics[width=0.9\textwidth]{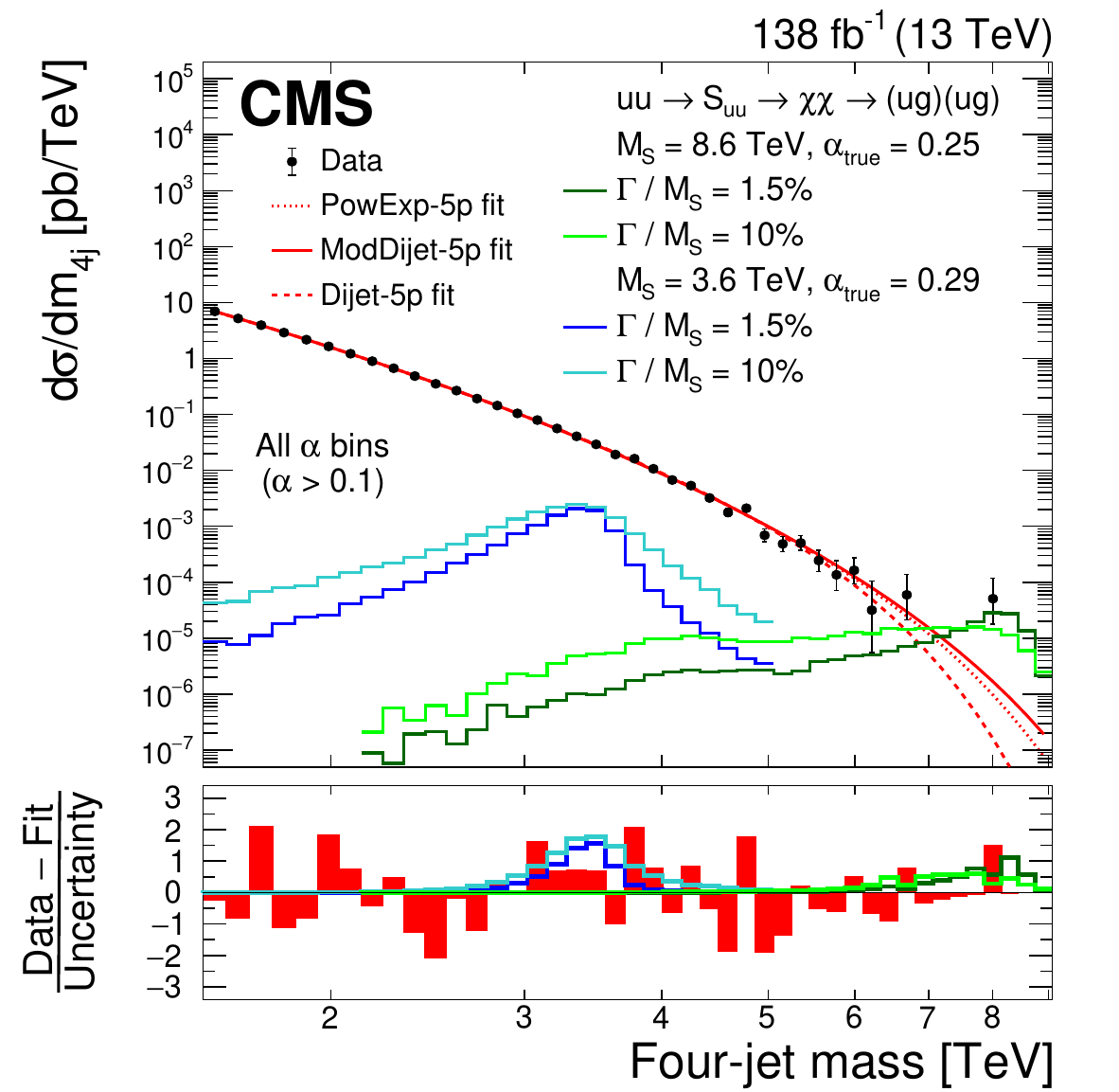}
\caption{The \mjjjj distribution in data (points), for all $\alpha$ bins combined, fitted with the background-only function, ModDijet-5p (red solid), and two alternative background functions, PowExp-5p and Dijet-5p (red dotted and dashed), with five free parameters. Examples of predicted diquark resonances with $\atrue = 0.25$, $M_{\PS} = 8.6\TeV$ and $\atrue = 0.29$, $M_{\PS} = 3.6\TeV$ are shown, with widths equal to 1.5 and 10\%. Their cross sections are equal to the observed upper limits at 95\% \CL. The lower panel shows the pulls from the fit with the background-only function to data, calculated using the statistical uncertainty of the data.} 
\label{fig:BkgFitsWithSignalsInclusive} 
\end{figure}

To better visualize how the broadest resonance considered more effectively encapsulates the three candidate events simultaneously, the inclusive \mjjjj distribution corresponding to the sum of all thirteen $\alpha$ bins is shown in Fig.~\ref{fig:BkgFitsWithSignalsInclusive}, along with the inclusive signal shapes. Figure~\ref{fig:BkgFitsWithSignalsInclusive} also shows the background-only fits with five parameter functional forms, identical to those of Ref.~\cite{CMS-EXO-21-010}.

As can be seen in Figs.~\ref{fig:BkgFitsWithSignalsAlphaBins_8p6} and~\ref{fig:BkgFitsWithSignalsInclusive}, broader resonances can result in signal events both at the resonance mass but also at lower \mjjjj. For example, for a 10\% broad resonance with $M_{\PS} = 8.6\TeV$ and $\atrue = 0.25$, 40\% of the signal is expected to be observed with $\mjjjj \leq 5.8\TeV$.

\section{Results}
\label{sec:Results}

Simulated signals and parametrized background shapes are fit to data, accounting for statistical and systematic uncertainties. As discussed, the data are the \mjjjj distributions in thirteen bins of $\alpha$, and the fit is performed simultaneously for all these bins. Upper limits at 95\% \CL are presented on the product of the signal cross section, branching fraction, and acceptance for all experimentally accessible values of resonance mass and width. 

For the signal extraction, we use a multibin counting-experiment likelihood, which is a product of Poisson distributions of the observed events in each bin. The likelihood is evaluated independently for each $M_{\PY}$ from 2 to 10\TeV in steps of 0.1\TeV, for $\Gamma/M_{\PY}$ values of 1.5, 5, and 10\%. The step sizes in resonance mass are comparable to the experimental mass resolution at the lower edge of each resonance mass interval. 

The systematic uncertainties considered in this analysis, and their handling, are identical to those presented in Ref.~\cite{CMS-EXO-21-010}. The sources of systematic uncertainty considered (values used), are the background , detailed below, the jet energy scale ($\pm$2\%), mass resolution of the four-jet and dijet resonances ($\pm$10\%) and the integrated luminosity ($\pm$1.6\%). These uncertainties are implemented as nuisance parameters in the likelihood model, with Gaussian constraints for the jet energy scale and resolution, and log-normal constraints for the integrated luminosity. The main source of the uncertainty in the background modeling comes from the choice of the background function and the range of possible parameter values of the background function. The parameters of the empirical functional form used to describe the background are considered as freely floating nuisances, and are evaluated via profiling. The discrete profiling method~\cite{Dauncey:2014xga} is used for considering the choice of the functional form as a discrete nuisance parameter, which is profiled in an analogous way to continuous nuisance parameters.

Next, we proceed to set limits and estimate significances using the \textsc{Combine} statistical package~\cite{Combine}. The modified frequentist criterion~\cite{Junk1999,bib-cls} is used to set upper limits on the signal cross sections, following the prescription described in Refs.~\cite{ATLAS:1379837,Cowan:2010js}, using the asymptotic approximation of the test statistic. In cases where a small number of events would make this approximation less accurate, the limit and significance estimation with respect to the background-only hypothesis is performed using test statistic distributions derived from pseudo-experiments.

\begin{figure}[hbtp]
\centering
\includegraphics[width=0.49\textwidth]{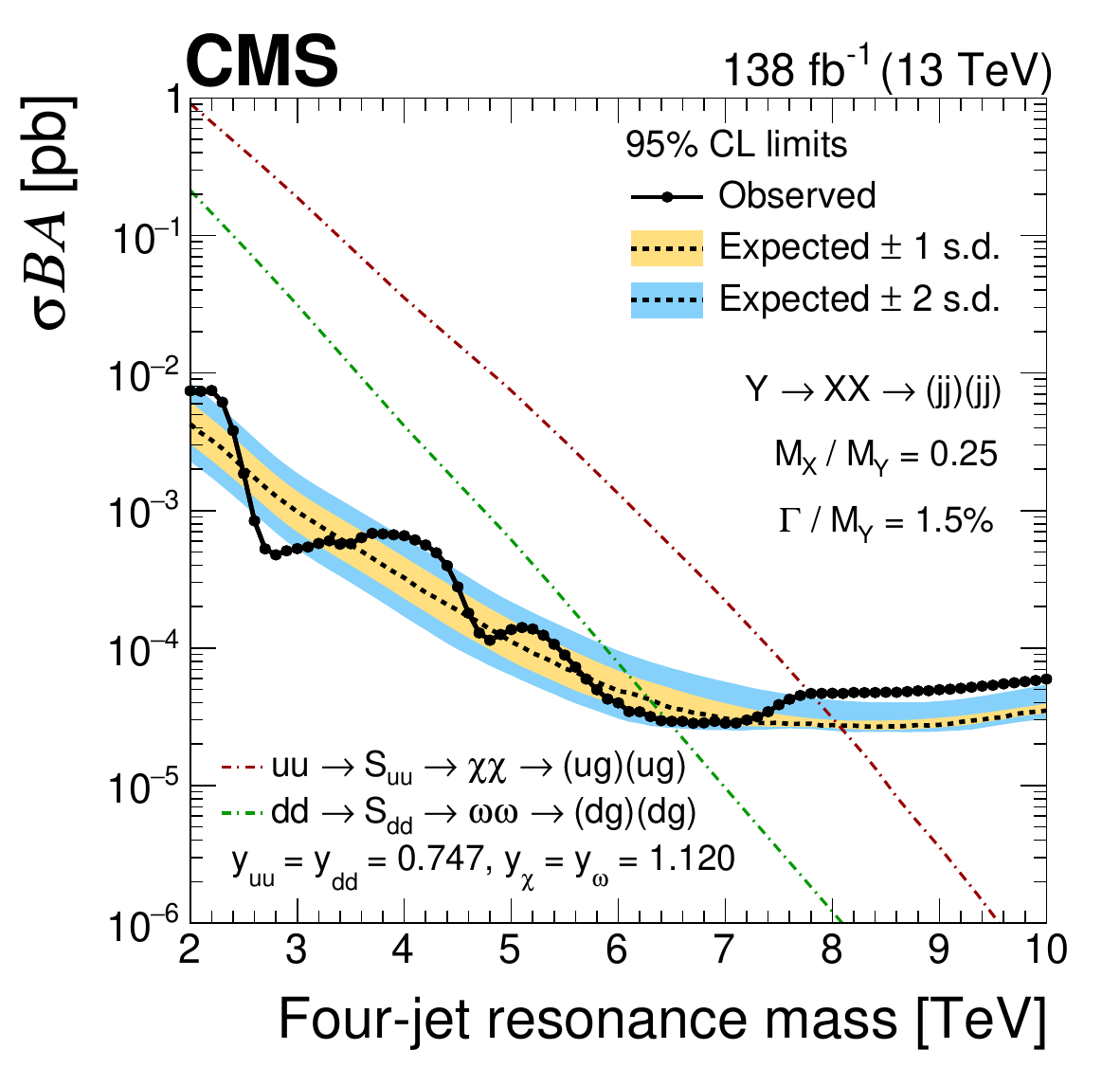}
\includegraphics[width=0.49\textwidth]{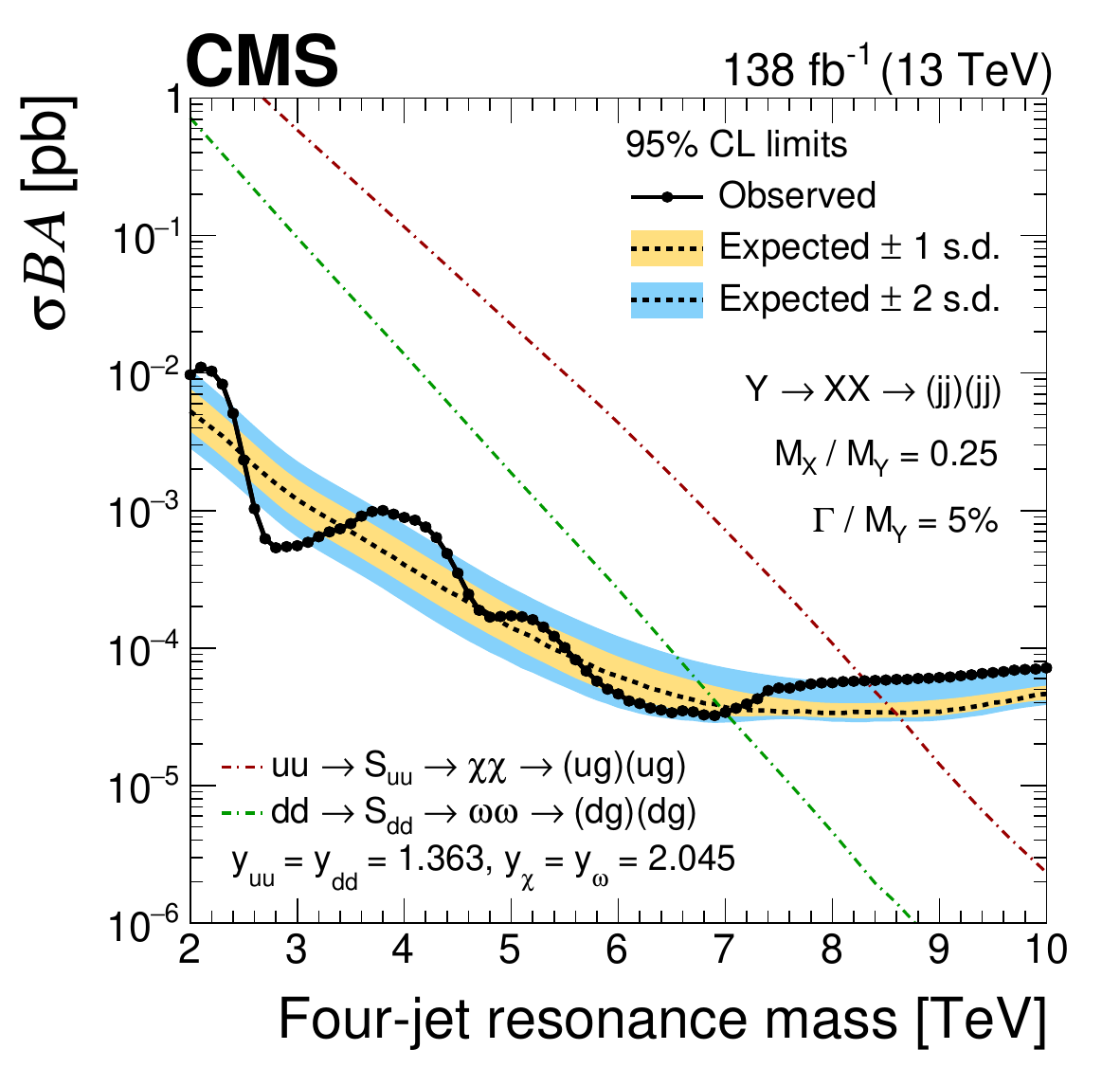}\\
\includegraphics[width=0.49\textwidth]{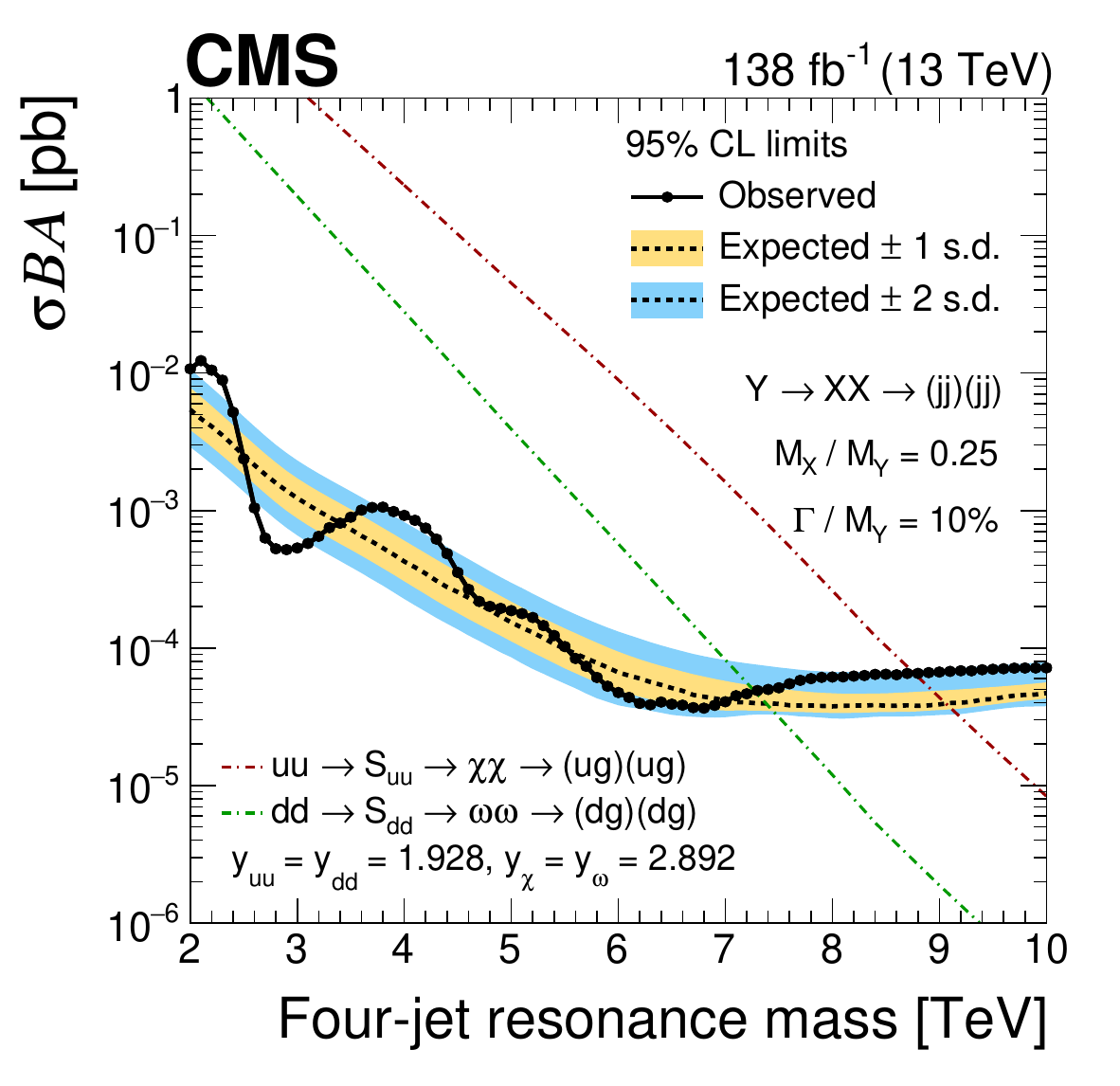}
\caption{The observed 95\% \CL upper limits (points) on the product of the cross section, branching fraction, and acceptance for resonant production of paired dijet resonances, \PX, with $\atrue = M_{\PX} / M_{\PY} = 0.25$ and width of the initial resonance \PY equal to 1.5 (upper left), 5 (upper right) and 10\% (lower). The corresponding expected limits (dashed) and their variations at the 1 and 2 s.d. levels (shaded bands) are also shown. Limits are compared to predictions for scalar \PSuu and \PSdd diquarks~\cite{Robertpaper} (dot-dashed lines) with couplings to pairs of up and down quarks, \yuu and \ydd, and to pairs of vector-like quarks, \ychi and \yomega, set appropriately in order to generate the corresponding widths.} 
\label{fig:LimitVsMass_Alpha0p25} 
\end{figure}

\begin{figure}[hbtp]
\centering
\includegraphics[width=0.9\textwidth]{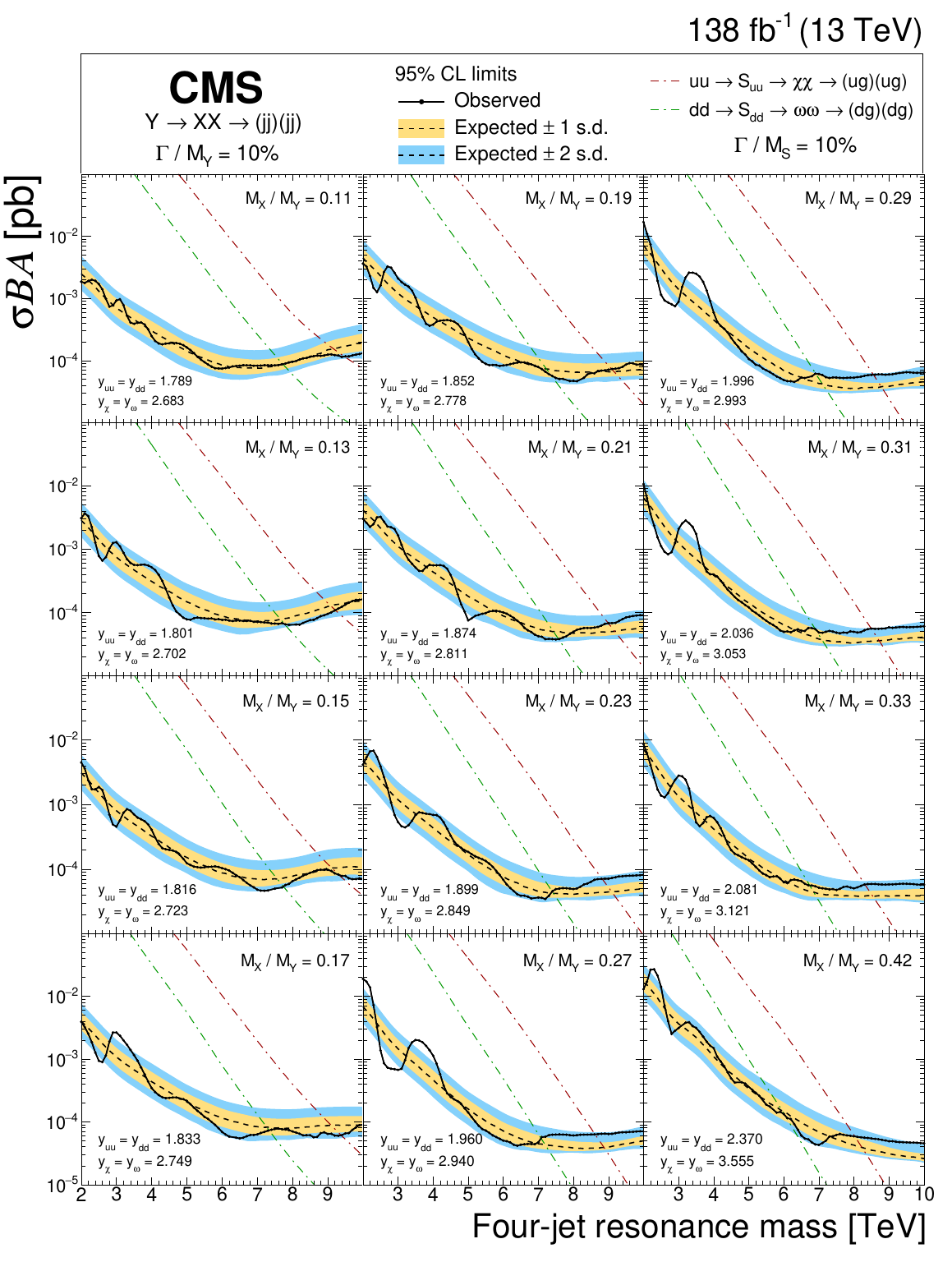}
\caption{The observed 95\% \CL upper limits (points) on the product of the cross section, branching fraction, and acceptance for resonant production of paired dijet resonances, \PX, with width of the initial resonance \PY equal to 10\%, and values of $\atrue = M_{\PX} / M_{\PY}$ shown in each panel. The corresponding expected limits (dashed) and their variations at the 1 and 2 s.d. levels (shaded bands) are also shown. Limits are compared to predictions for scalar \PSuu and \PSdd diquarks~\cite{Robertpaper} (dot-dashed lines) with couplings to pairs of up and down quarks, \yuu and \ydd, and to pairs of vector-like quarks, \ychi and \yomega, set appropriately in order to generate a 10\% width.} 
\label{fig:LimitVsMass_AllAlpha_W0p1} 
\end{figure}

\begin{figure}[hbtp]
\centering
\includegraphics[width=0.9\textwidth]{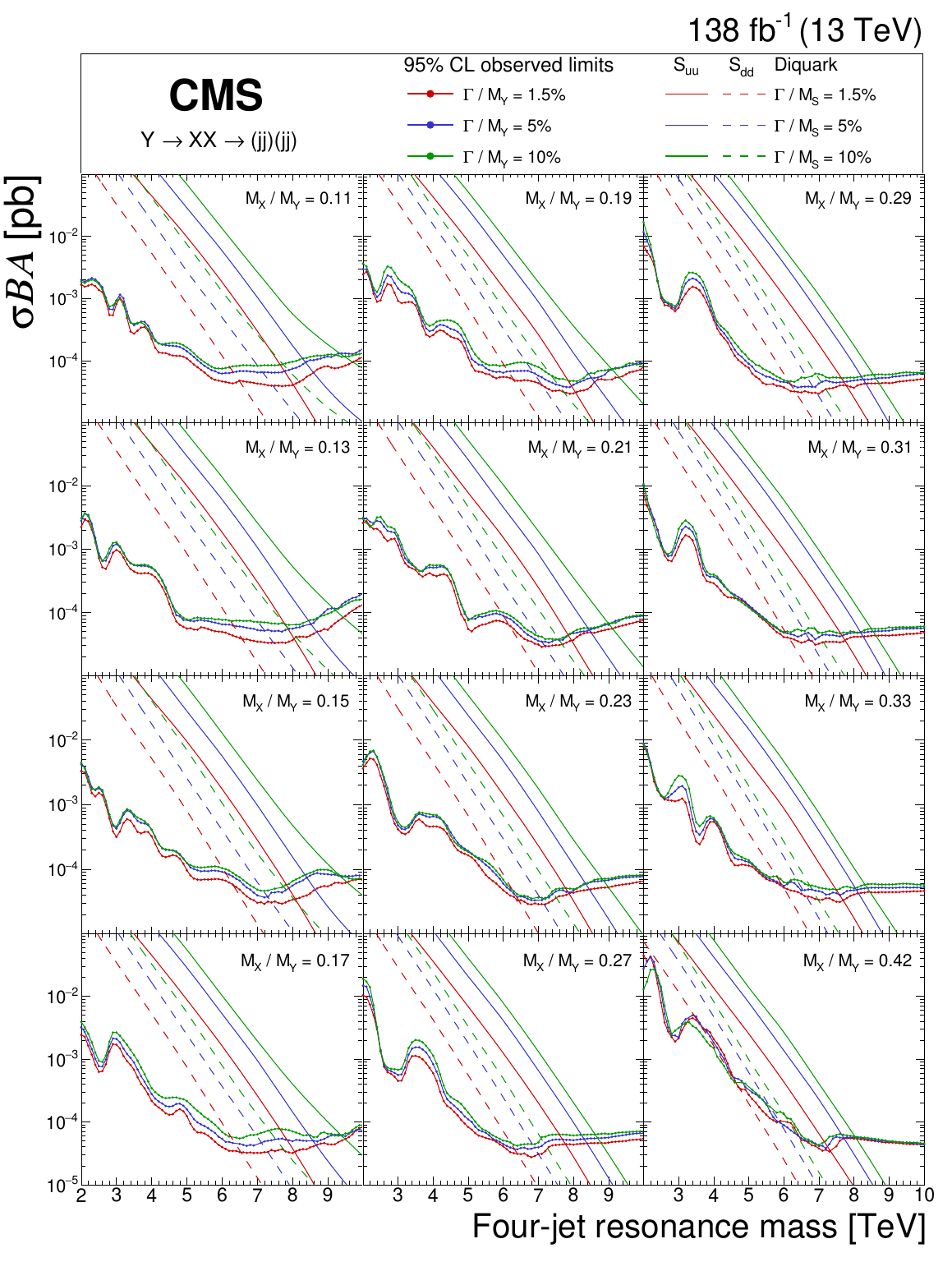}
\caption{The observed 95\% \CL upper limits (points) on the product of the cross section, branching fraction, and acceptance for resonant production of paired dijet resonances, \PX, with the values of $\atrue = M_{\PX} / M_{\PY}$ shown in each panel. Different colors correspond to the various widths of the \PY resonance. Limits are compared to predictions for scalar \PSuu (solid lines) and \PSdd (dashed lines) diquarks~\cite{Robertpaper} with couplings to pairs of up and down quarks, \yuu and \ydd, and to pairs of vector-like quarks, \ychi and \yomega, set appropriately in order to generate the corresponding widths.} 
\label{fig:ObsLimitVsMass_AllAlpha} 
\end{figure}

The upper limits at 95\% \CL obtained from this procedure are presented in Figs.~\ref{fig:LimitVsMass_Alpha0p25}--\ref{fig:ObsLimitVsMass_AllAlpha}. First, we show in Fig.~\ref{fig:LimitVsMass_Alpha0p25} limits for the three considered widths, for signals with $\atrue = 0.25$, the value of \atrue where the maximum significance in the 8--10\TeV resonance mass range is achieved. Figure~\ref{fig:LimitVsMass_Alpha0p25} presents the model-independent observed and expected upper limits on $\sigma \mathcal{B} A$, representing the product of the cross section ($\sigma$), branching fraction ($\mathcal{B}$), and acceptance ($A$) for the kinematic requirements. Limits are quoted within the acceptance, as presented in Fig.~\ref{fig:AcceptanceTimesEfficiency}, and corrected for the efficiency, which can be inferred from Fig.~\ref{fig:AcceptanceTimesEfficiency}. In Fig.~\ref{fig:LimitVsMass_Alpha0p25}, the observed limit is greater than the expected limit at resonance masses of around 2, 4, and 8\TeV, and less than the expected limit near a resonance mass of 3\TeV, due to similar effects in the pulls near these values of \mjjjj in Fig.~\ref{fig:BkgFitsWithSignalsInclusive}. The observed and expected limits on resonant production for the largest width considered in the analysis are shown in Fig.~\ref{fig:LimitVsMass_AllAlpha_W0p1}, for twelve values of \atrue chosen at the center of each bin of $\alpha$. In Fig.~\ref{fig:ObsLimitVsMass_AllAlpha}, the observed limits are shown for all \atrue values and are superimposed for each width considered.

Mass limits on new particles can be determined from comparisons of the presented upper limits on $\sigma \mathcal{B} A$ with the corresponding background predictions. In Figs.~\ref{fig:LimitVsMass_Alpha0p25}--\ref{fig:ObsLimitVsMass_AllAlpha}, we compare these limits to the predicted leading order $\sigma \mathcal{B}$ for the scalar \PSuu and \PSdd diquark models, multiplied by the $A$ in Fig.~\ref{fig:AcceptanceTimesEfficiency}. The cross section from this prediction is not proportional to the width, but rather, it is larger than such proportionality predicts at large resonance masses, due to the dominance of the long, low-mass tails. Since these do not contribute to the signal search, as they lie in a region that is dominated by background, we adjust the cross section from the theoretical prediction using an efficiency factor that isolates the \mjjjj range where the expected significance exceeds 99\% of that obtained without the correction. 

For a given model, new particles are excluded at 95\% \CL in mass regions where the theoretical prediction lies at or above the observed upper limit in Figs.~\ref{fig:LimitVsMass_Alpha0p25}--\ref{fig:ObsLimitVsMass_AllAlpha}. For example, for \PSuu and \PSdd signals with $\atrue = 0.25$, exclusion mass limits are derived from Fig.~\ref{fig:LimitVsMass_Alpha0p25} and presented in Table~\ref{tab:mass_limits}. Note that in Table~\ref{tab:mass_limits}, for the 8.6\TeV resonance with a width of 10\%, while the benchmark \PSuu diquark model is excluded, the \PSdd diquark model is not excluded and remains a viable hypothesis for our three candidate events.

\begin{table}[htb]
\centering
\topcaption{Observed and expected mass limits at 95\% \CL for \PSuu and \PSdd diquark models with $\atrue = 0.25$.}
\label{tab:mass_limits}
\renewcommand{\arraystretch}{1.1}
\begin{tabular}{lcc} 
\multirow{2}*{Width}                         & \multicolumn{2}{c}{Observed (expected) mass limit [$\TeVns$]}  \\
                                                         & \PSuu scalar diquark                  & \PSdd scalar diquark  \\
\hline                               
$\Gamma / M_{\PS} = 1.5\%$  & 7.8 (8.1) &  6.5 (6.3)     \\   
$\Gamma / M_{\PS} = 5\%$    & 8.3 (8.6) &  7.0 (6.9)     \\   
$\Gamma / M_{\PS} = 10\%$   & 8.8 (9.1) &  7.3 (7.4)     \\    
\end{tabular}
\end{table}

\begin{figure}[hbtp]
\centering
\includegraphics[width=0.49\textwidth]{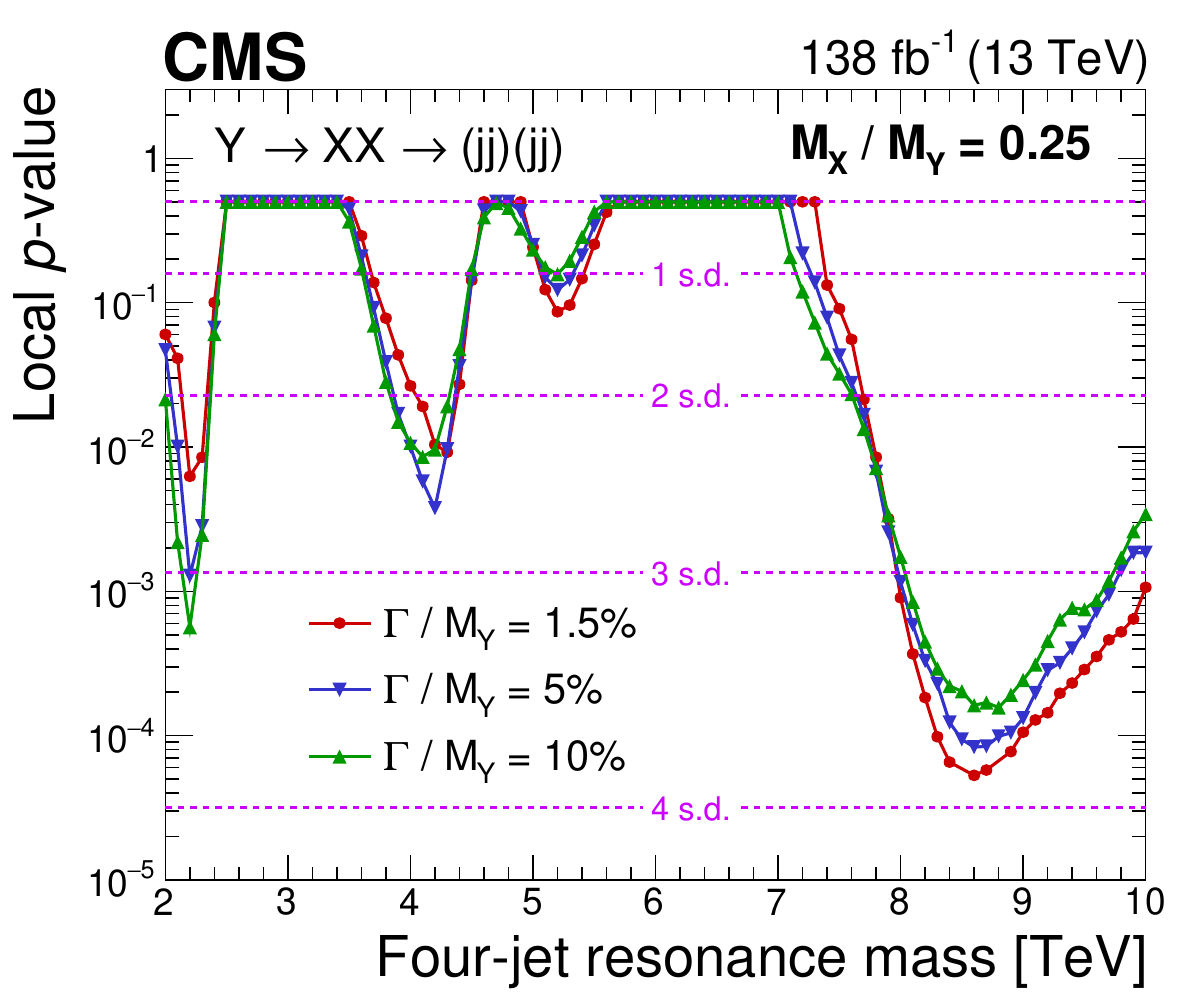} 
\includegraphics[width=0.49\textwidth]{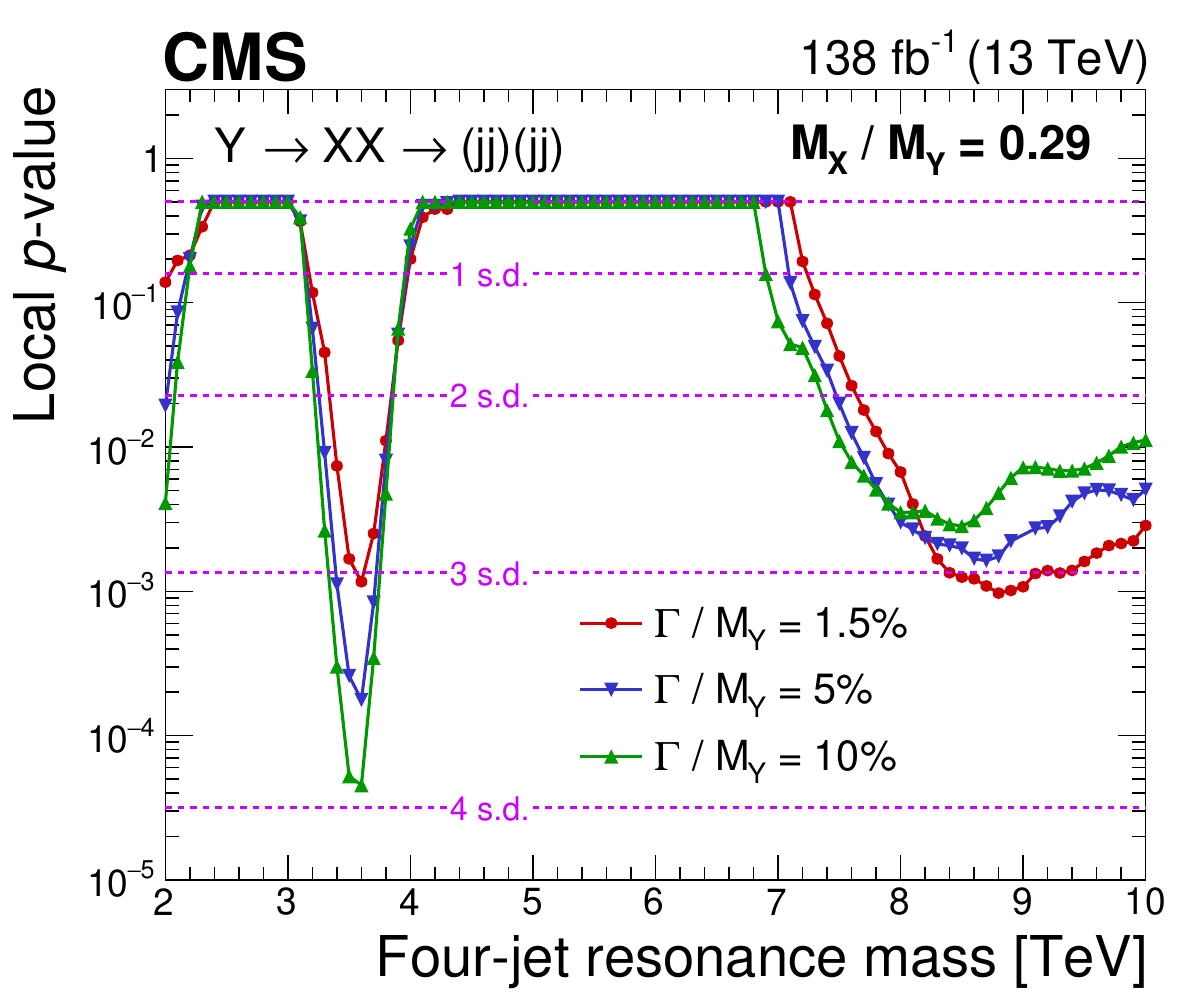}
\caption{Observed local $p$-value for a four-jet resonance, \PY, decaying to a pair of dijet resonances, \PX, with $\atrue = M_{\PX} / M_{\PY} = 0.25$ (\cmsLeft) and 0.29 (\cmsRight), and various widths of \PY superimposed. Also shown are corresponding levels of local significance (dashed lines) in units of s.d.} 
\label{fig:PValueVsMassVsWidth} 
\end{figure}

The local $p$-value of a signal, as a function of four-jet resonance mass and width, is presented for $\atrue = 0.25$ in Fig.~\ref{fig:PValueVsMassVsWidth} (\cmsLeft). 
For a width of 1.5\%, the most significant signal hypothesis occurs at a four-jet resonance mass of 8.6\TeV and a dijet resonance mass of 2.15\TeV, for which the local significance is 3.9 s.d, the same as the significance observed for a narrow resonance with those mass values in Ref.~\cite{CMS-EXO-21-010}. As the four-jet resonance becomes broader, Fig.~\ref{fig:PValueVsMassVsWidth} demonstrates that the local significance remains high, exceeding or being equal to 3.6 s.d., even for the largest width of 10\%. Hence, the significance of an 8.6\TeV four-jet resonance is found to be relatively insensitive to the choice of width, making a broad resonance an equally valid interpretation of the excess. For the look-elsewhere-effect~\cite{Gross_2010}, a large number of possible signals in the 2D plane of four-jet resonance mass and \atrue are considered, resulting in a global significance of 1.6 s.d. for a narrow resonance and 1.4 s.d. for the broadest resonance considered. 
 
Table~\ref{tab:best_fit_r_values} summarizes these local and global significances for a four-jet resonance, along with the best fit value of $\sigma \mathcal{B} A$ and number of events, as a function of width for a four-jet resonance mass of 8.6\TeV with $\atrue = 0.25$. The table demonstrates that broader resonances, with widths of 5 and 10\%, provide a better fit for all three observed candidate events.

To explore all resonance masses, the entire 2D plane of four-jet resonance mass and \atrue is scanned. A second excess is observed at a four-jet resonance mass of 3.6\TeV, $\atrue = 0.29$, and 10\% width with a local and global significance of 3.9 and 2.2 s.d. respectively, as shown in Fig.~\ref{fig:PValueVsMassVsWidth} (\cmsRight). This excess points to a structure in data that was also reported by the search for nonresonant production of pairs of dijet resonances in Ref.~\cite{CMS-EXO-21-010}, and further discussed in Ref.~\cite{Crivellin:2022nms}. The distributions shown in the three $\alpha$ bins of Fig.~\ref{fig:BkgFitsWithSignalsAlphaBins_3p6} allow comparisons to be made between the observed and predicted \mjjjj distributions in the region of this excess. Its corresponding best fit values of $\sigma \mathcal{B} A$, number of events, and signal significance are presented in Table~\ref{tab:best_fit_r_values}. Here the significance increases with width.

\begin{table}[htbp]
\centering
\topcaption{For each resonance width, for the two most significant resonance hypotheses found, we list the \PY and \PX resonance masses, the best fit values of signal $\sigma \mathcal{B} A$ from the signal plus background fit, the corresponding number of signal events, and the local and global significance.}
\label{tab:best_fit_r_values}
\renewcommand{\arraystretch}{1.3}
\begin{tabular}{lccccc} 
$\Gamma / M_{\PY}$  &   $M_{\PY}$  &  $M_{\PX}$   &    $\sigma \mathcal{B} A$  &   Events  &  Local (global) significance  \\[-1.ex] 
$[\%]$  &  $[\TeVns]$  & $[\TeVns]$   &  [pb]  &    &  [s.d.]  \\
\hline                              
1.5  & 8.6 & 2.15 &  $1.68^{+1.50}_{-0.94} \times 10^{-5}$ & $2.3^{+2.1}_{-1.3}$   &  3.9 (1.6) \\ 
5    &  "  &  "   &  $2.04^{+1.83}_{-1.15} \times 10^{-5}$ & $2.8^{+2.5}_{-1.6}$   &  3.8 (1.5) \\ 
10   &  "  &  "   &  $2.22^{+1.99}_{-1.26} \times 10^{-5}$ & $3.1^{+2.7}_{-1.7}$   &  3.6 (1.4) \\
1.5  & 3.6 &  1.0 &  $0.83^{+0.30}_{-0.32} \times 10^{-3}$ & $114^{+42}_{-44}$     &  3.0 (0.5) \\ 
5    &  "  &  "   &  $1.18^{+0.36}_{-0.33} \times 10^{-3}$ & $162^{+50}_{-46}$     &  3.6 (1.6) \\ 
10   &  "  &  "   &  $1.53^{+0.42}_{-0.40} \times 10^{-3}$ & $210^{+58}_{-54}$     &  3.9 (2.2) \\
\end{tabular}
\end{table}

\section{The third candidate event}
\label{sec:CandidateEvent}

The high significance of narrow resonances at 8.6\TeV is predominantly caused by the two events with $\mjjjj = 8\TeV$ and $\mjj = 2\TeV$, shown in Fig.~\ref{fig:2D_Data_QCD}. Although similarly isolated from the bulk of the 2D mass distribution, the third event, with $\mjjjj = 5.8\TeV$ and $\mjj = 2.0\TeV$, has minimal contribution in the fit with narrow resonances. However, as Fig.~\ref{fig:2D_Data_QCD} and Table~\ref{tab:best_fit_r_values} indicate, it is notably more compatible with broad resonances.   

\begin{figure}[bp]
\centering
\includegraphics[width=0.95\textwidth]{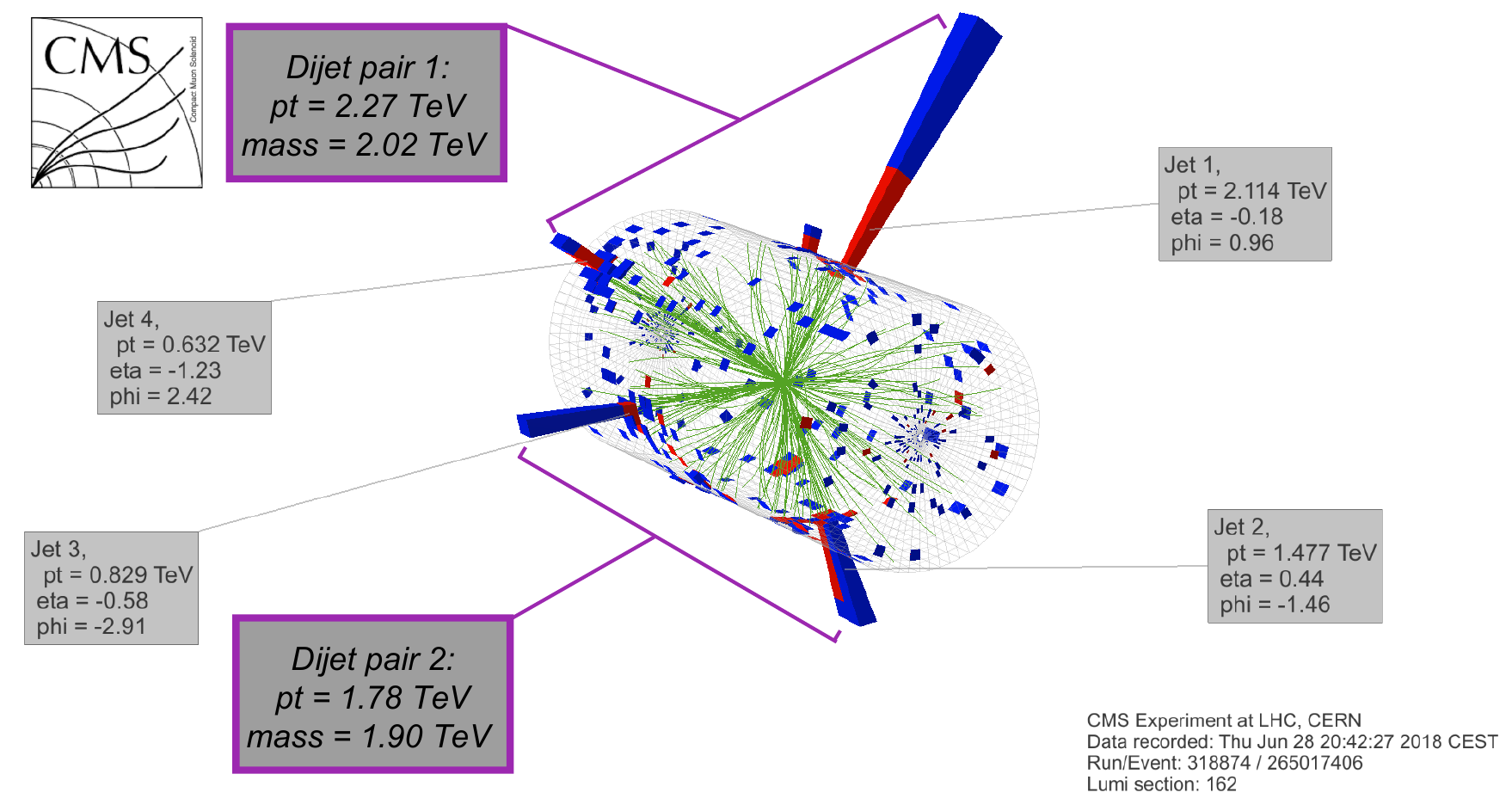}
\caption{The 3D display of the candidate event for broad resonances with a four-jet mass of 5.8\TeV. The display shows the energy deposited in the electromagnetic (red) and hadronic (blue) calorimeters and the reconstructed tracks of charged particles (green). The grouping of the four observed jets into two dijet pairs (purple boxes) is discussed in the text.} 
\label{fig:3DEventDisplay} 
\end{figure}

The 3D display of this event is shown in Fig.~\ref{fig:3DEventDisplay}. It exhibits very similar topology and characteristics to the two events with $\mjjjj = 8\TeV$ that were shown in Ref.~\cite{CMS-EXO-21-010}. It was collected during 2018 and has $\mjjjj = 5.8\TeV$. The invariant mass of the dijet pair that includes jet 1 and jet 4 is 2.0\TeV, and the mass of the other pair, from jet 2 and jet 3, is 1.9\TeV. The two dijet pairs are back-to-back in azimuthal angle ($\abs{\Delta\phi} = 3.1$), nearby in pseudorapidity ($\abs{\Delta\eta} = 0.7$), and have a mass asymmetry of 0.03. The separation in $\eta$-$\phi$ space between the two jets within each dijet pair is 1.8, the same value for both dijet pairs. The \pt , $\eta$, and $\phi$ of each jet can be seen in Fig.~\ref{fig:3DEventDisplay}.

\section{Summary}
\label{sec:Summary}

A search for broad resonances, decaying to pairs of dijet resonances with equal masses, has been presented. The results are a reinterpretation of a similar narrow resonance search~\cite{CMS-EXO-21-010}, using data corresponding to an integrated luminosity of 138\fbinv, from proton-proton collisions at $\sqrt{s}=13\TeV$.

Empirical functions that model the background, and simulated shapes of resonance signals with widths of 1.5, 5, and 10\%, are fit to the observed four-jet mass (\mjjjj) distributions, separated into bins of the ratio of the dijet to four-jet distributions. There are three events in the tails of the distributions, two with an \mjjjj of 8\TeV and one with an \mjjjj of 5.8\TeV, all of which have an average dijet mass of approximately 2\TeV, that result in an excess. Although the event with an \mjjjj of 5.8\TeV contributes minimally to a fit with narrow resonances, it is far more compatible with broad resonances of the same mass. Hence, the local significance for a four-jet resonance remains above 3.6 standard deviations (s.d.) even for the largest width considered, leading to the conclusion that broad resonances are an equally valid interpretation of the excess. An event reported by the ATLAS Collaboration~\cite{PairedDijetATLAS}, with an \mjjjj of 6.6\TeV and an \mjj of 2.2\TeV, falls within the CMS 68\% probability contour for a 5 or 10\% broad resonance with a mass of 8.4\TeV, and hence is likely compatible with those two hypotheses.

A second excess with a local (global) significance of 3.9 (2.2) s.d. is observed at a four-jet resonance mass of 3.6\TeV, with an average dijet resonance mass of 1.0\TeV, for a mediator width of 10\%. This excess was previously reported with local and global significances of 3.6 and 2.5 s.d., respectively, in Ref.~\cite{CMS-EXO-21-010}, originating from the dijet data near a mass of 1\TeV, in the search for nonresonant production of pairs of dijet resonances.

Model-independent upper limits at 95\% confidence level are presented on the product of the cross section, branching fraction and acceptance as a function of the four-jet resonance mass between 2 and 10\TeV, for all accessible values of the ratio of the dijet to four-jet resonance masses, for widths ranging from 1.5 to 10\%. Limits are compared to models~\cite{Robertpaper} of diquarks, which decay to pairs of vector-like quarks, which in turn decay to a quark and a gluon. Mass limits are presented for all accessible values of the ratio of the vector-like quark to diquark masses, for the three values of diquark width considered.

\begin{acknowledgments}
We congratulate our colleagues in the CERN accelerator departments for the excellent performance of the LHC and thank the technical and administrative staffs at CERN and at other CMS institutes for their contributions to the success of the CMS effort. In addition, we gratefully acknowledge the computing centers and personnel of the Worldwide LHC Computing Grid and other centers for delivering so effectively the computing infrastructure essential to our analyses. Finally, we acknowledge the enduring support for the construction and operation of the LHC, the CMS detector, and the supporting computing infrastructure provided by the following funding agencies: SC (Armenia), BMBWF and FWF (Austria); FNRS and FWO (Belgium); CNPq, CAPES, FAPERJ, FAPERGS, and FAPESP (Brazil); MES and BNSF (Bulgaria); CERN; CAS, MoST, and NSFC (China); MINCIENCIAS (Colombia); MSES and CSF (Croatia); RIF (Cyprus); SENESCYT (Ecuador); ERC PRG, RVTT3 and MoER TK202 (Estonia); Academy of Finland, MEC, and HIP (Finland); CEA and CNRS/IN2P3 (France); SRNSF (Georgia); BMBF, DFG, and HGF (Germany); GSRI (Greece); NKFIH (Hungary); DAE and DST (India); IPM (Iran); SFI (Ireland); INFN (Italy); MSIT and NRF (Republic of Korea); MES (Latvia); LMTLT (Lithuania); MOE and UM (Malaysia); BUAP, CINVESTAV, CONACYT, LNS, SEP, and UASLP-FAI (Mexico); MOS (Montenegro); MBIE (New Zealand); PAEC (Pakistan); MES and NSC (Poland); FCT (Portugal); J MESTD (Serbia); MICIU/AEI and PCTI (Spain); MOSTR (Sri Lanka); Swiss Funding Agencies (Switzerland); MST (Taipei); MHESI and NSTDA (Thailand); TUBITAK and TENMAK (T\"{u}rkiye); NASU (Ukraine); STFC (United Kingdom); DOE and NSF (USA).

\hyphenation{Rachada-pisek} Individuals have received support from the Marie-Curie program and the European Research Council and Horizon 2020 Grant, contract Nos.\ 675440, 724704, 752730, 758316, 765710, 824093, 101115353, 101002207, 101001205, and COST Action CA16108 (European Union); the Leventis Foundation; the Alfred P.\ Sloan Foundation; the Alexander von Humboldt Foundation; the Science Committee, project no. 22rl-037 (Armenia); the Fonds pour la Formation \`a la Recherche dans l'Industrie et dans l'Agriculture (FRIA-Belgium); the Beijing Municipal Science \& Technology Commission, No. Z191100007219010, the Fundamental Research Funds for the Central Universities, the Ministry of Science and Technology of China under Grant No. 2023YFA1605804, and the Natural Science Foundation of China under Grant No. 12061141002 (China); the Ministry of Education, Youth and Sports (MEYS) of the Czech Republic; the Shota Rustaveli National Science Foundation, grant FR-22-985 (Georgia); the Deutsche Forschungsgemeinschaft (DFG), among others, under Germany's Excellence Strategy -- EXC 2121 ``Quantum Universe" -- 390833306, and under project number 400140256 - GRK2497; the Hellenic Foundation for Research and Innovation (HFRI), Project Number 2288 (Greece); the Hungarian Academy of Sciences, the New National Excellence Program - \'UNKP, the NKFIH research grants K 131991, K 133046, K 138136, K 143460, K 143477, K 146913, K 146914, K 147048, 2020-2.2.1-ED-2021-00181, TKP2021-NKTA-64, and 2021-4.1.2-NEMZ\_KI-2024-00036 (Hungary); the Council of Science and Industrial Research, India; ICSC -- National Research Center for High Performance Computing, Big Data and Quantum Computing, FAIR -- Future Artificial Intelligence Research, and CUP I53D23001070006 (Mission 4 Component 1), funded by the NextGenerationEU program (Italy); the Latvian Council of Science; the Ministry of Education and Science, project no. 2022/WK/14, and the National Science Center, contracts Opus 2021/41/B/ST2/01369 and 2021/43/B/ST2/01552 (Poland); the Funda\c{c}\~ao para a Ci\^encia e a Tecnologia, grant CEECIND/01334/2018 (Portugal); the National Priorities Research Program by Qatar National Research Fund; MICIU/AEI/10.13039/501100011033, ERDF/EU, "European Union NextGenerationEU/PRTR", and Programa Severo Ochoa del Principado de Asturias (Spain); the Chulalongkorn Academic into Its 2nd Century Project Advancement Project, and the National Science, Research and Innovation Fund via the Program Management Unit for Human Resources \& Institutional Development, Research and Innovation, grant B39G680009 (Thailand); the Kavli Foundation; the Nvidia Corporation; the SuperMicro Corporation; the Welch Foundation, contract C-1845; and the Weston Havens Foundation (USA).
\end{acknowledgments}

\bibliography{auto_generated} 
\cleardoublepage \appendix\section{The CMS Collaboration \label{app:collab}}\begin{sloppypar}\hyphenpenalty=5000\widowpenalty=500\clubpenalty=5000
\cmsinstitute{Yerevan Physics Institute, Yerevan, Armenia}
{\tolerance=6000
A.~Hayrapetyan, V.~Makarenko\cmsorcid{0000-0002-8406-8605}, A.~Tumasyan\cmsAuthorMark{1}\cmsorcid{0009-0000-0684-6742}
\par}
\cmsinstitute{Institut f\"{u}r Hochenergiephysik, Vienna, Austria}
{\tolerance=6000
W.~Adam\cmsorcid{0000-0001-9099-4341}, J.W.~Andrejkovic, L.~Benato\cmsorcid{0000-0001-5135-7489}, T.~Bergauer\cmsorcid{0000-0002-5786-0293}, M.~Dragicevic\cmsorcid{0000-0003-1967-6783}, C.~Giordano, P.S.~Hussain\cmsorcid{0000-0002-4825-5278}, M.~Jeitler\cmsAuthorMark{2}\cmsorcid{0000-0002-5141-9560}, N.~Krammer\cmsorcid{0000-0002-0548-0985}, A.~Li\cmsorcid{0000-0002-4547-116X}, D.~Liko\cmsorcid{0000-0002-3380-473X}, M.~Matthewman, I.~Mikulec\cmsorcid{0000-0003-0385-2746}, J.~Schieck\cmsAuthorMark{2}\cmsorcid{0000-0002-1058-8093}, R.~Sch\"{o}fbeck\cmsAuthorMark{2}\cmsorcid{0000-0002-2332-8784}, D.~Schwarz\cmsorcid{0000-0002-3821-7331}, M.~Shooshtari, M.~Sonawane\cmsorcid{0000-0003-0510-7010}, W.~Waltenberger\cmsorcid{0000-0002-6215-7228}, C.-E.~Wulz\cmsAuthorMark{2}\cmsorcid{0000-0001-9226-5812}
% ...
\par}
\vskip\cmsinstskip
\dag:~Deceased\\
$^{1}$Also at Yerevan State University, Yerevan, Armenia\\
$^{2}$Also at TU Wien, Vienna, Austria\\
% ...
\end{sloppypar}
%%% END EDITABLE REGION %%%
% skeleton_end
\end{document}